\documentclass[journal]{IEEEtran}

\usepackage{cite}
\usepackage{amsmath,epsfig,amssymb}
\usepackage[ruled,vlined]{algorithm2e}
\usepackage{psfrag}
\usepackage{subfigure}
  \usepackage{float}
  \usepackage{cases}
\usepackage[amsmath,thmmarks]{ntheorem}
\usepackage{color}
\usepackage{tikz}
\usepackage{graphicx}

\usepackage{booktabs,caption}
\usepackage[flushleft]{threeparttable}

\usepackage{verbatim} 
\usepackage{glossaries}

\usepackage{stfloats, flushend}

 \floatstyle{ruled}
 \newfloat{algorithm}{htb}{loa}
 \floatname{algorithm}{\textbf{Algorithm}}
 \makeatletter
 \providecommand*{\toclevel@algorithm}{0} 
 \makeatother

\theoremsymbol{$\Box$}
\newtheorem{remark}{Remark}

\theoremsymbol{$\Box$}
\newtheorem{definition}{Definition}

\theoremsymbol{$\Box$}
\newtheorem{lemma}{Lemma}

\theoremsymbol{$\Box$}

\theoremsymbol{$\Box$}
\newtheorem{proposition}{Proposition}

\theoremsymbol{$\Box$}
\newtheorem{corollary}{Corollary}

\theoremsymbol{$\Box$}

\theoremsymbol{$\Box$}

\theoremsymbol{$\Box$}

\makeatletter
\newcommand\unmarkfootnote[1]{%
  \begingroup
    \let\@makefntext\noindent
    \footnotetext{#1}%
  \endgroup}
\makeatother

\makeatletter

\newcommand{\Rmnum}[1]{\expandafter\@slowromancap\romannumeral #1@}
\makeatother

\newcommand{\mat}{\boldsymbol}

\usepackage{footnote}

\usepackage{xspace}
\usepackage{setspace}

\newcommand{\SNR}{\ensuremath{\text{\textup{SNR}}}\xspace}

\newcommand{\diag}{\mathop{\mathrm{diag}}}
\newcommand{\rank}{\mathop{\mathrm{rank}}}
\newcommand{\trace}{\mathop{\mathrm{Tr}}}

\makeglossaries

\newacronym{mac}{MAC}{multiple-access channel}
\newacronym{bc}{BC}{broadcast channel}
\newacronym{mimo}{MIMO}{multiple-input multiple-output}
\newacronym{siso}{SISO}{single-input single-output}
\newacronym{af}{AF}{amplify-and-forward}
\newacronym{df}{DF}{decode-and-forward}
\newacronym{cf}{CF}{compress-and-forward}
\newacronym{mwrc}{MWRC}{multi-way relay channel}
\newacronym{pde}{PDE}{partial data exchange}
\newacronym{fde}{FDE}{full data exchange}
\newacronym{iid}{i.i.d.}{independent and identically distributed}
\newacronym{awgn}{AWGN}{additive white Gaussian noise}
\newacronym{awg}{AWG}{additive white Gaussian}
\newacronym{sic}{SIC}{successive interference cancellation}
\newacronym{snr}{SNR}{signal-to-noise ratio}
\newacronym{sinr}{SINR}{signal to interference plus noise ratio}
\newacronym{zf}{ZF}{zero-forcing}
\newacronym{mmse}{MMSE}{minimum mean square error}
\newacronym{sud}{SUD}{single user decoding}
\newacronym{dof}{DoF}{degrees of freedom}
\newacronym{gdof}{GDoF}{generalized degrees of freedom}
\newacronym{nnc}{NNC}{noisy network coding}
\newacronym{dmn}{DMN}{discrete memoryless network}
\newacronym{csi}{CSI}{channel state information}
\newacronym{ee}{EE}{energy efficiency}
\newacronym{ian}{IAN}{treating interference as noise}
\newacronym{snd}{SND}{simultaneous non-unique decoding}
\newacronym{brd}{BRD}{best response dynamics}
\newacronym{br}{BR}{best response}
\newacronym{ne}{NE}{Nash equilibrium}
\newacronym{lhs}{LHS}{left-hand side}
\newacronym{rhs}{RHS}{right-hand side}
\newacronym{ia}{IA}{interference alignment}

\begin{document}

\title{Grouping-based Interference Alignment with IA-Cell Assignment  in Multi-Cell MIMO MAC under Limited Feedback}
\author{Pan Cao, \IEEEmembership{Member, IEEE},
       Alessio Zappone, \IEEEmembership{Member, IEEE}, \\
        Eduard A. Jorswieck, \IEEEmembership{Senior Member, IEEE} 
\thanks{Copyright (c) 2015 IEEE. Personal use of this material is permitted. However, permission to use this material for any other purposes must be obtained from the IEEE by sending a request to pubs-permissions@ieee.org. 

The work of P. Cao is in part funded by the Engineering and Physical Sciences Research Council (EPSRC) project SERAN, under grant EP/L026147/1; The work of A. Zappone has been funded by the German Research Foundation (DFG) project CEMRIN, under grant ZA 747/1-3; The work of E. Jorswieck was supported in part by the German Research Foundation, Deutsche Forschungsgemeinschaft (DFG) in the Collaborative Research Center 912 Highly Adaptive Energy-Efficient Computing. 

P. Cao was with the Chair of Communications Theory, Communications Laboratory, TU Dresden, Dresden 01062, Germany. He is now with the Institute for Digital Communications, The University of Edinburgh, Edinburgh EH9 3JL, UK (e-mail: P. Cao@ed.ac.uk).

A. Zappone and E. Jorswieck are with the Chair of Communications Theory, Communications Laboratory, TU Dresden, Dresden 01062, Germany (e-mail: \{Alessio.Zappone, Eduard.Jorswieck\}@tu-dresden.de).}}

\maketitle

\begin{abstract}
Interference alignment (IA) is a promising technique to efficiently mitigate interference and to enhance the capacity of a wireless communication network. This paper proposes a grouping-based interference alignment (GIA) with optimized IA-Cell assignment for the multiple cells  interfering multiple-input multiple-output (MIMO) multiple access channel (MAC) network under limited feedback. 
This work consists of three main parts: 1) an improved version (including some new improvements) of the GIA with respect to the degrees of freedom (DoF) and optimal linear transceiver design is provided, which allows for low-complexity and distributed implementation; 2) based on the GIA, the concept of IA-Cell assignment is introduced. Three IA-Cell assignment algorithms are proposed with different backhaul overhead and their DoF and rate performance is investigated;
3) the performance of the proposed GIA algorithms is studied under limited feedback of IA precoders. To enable efficient feedback, a dynamic feedback bit allocation (DBA) problem is formulated and solved in closed-form. The practical implementation, the backhaul overhead requirements, and the complexity of the proposed algorithms are analyzed. Numerical results show that our proposed algorithms greatly outperform the traditional GIA under both unlimited and limited feedback.
\end{abstract}

\IEEEpeerreviewmaketitle

\section{Introduction}\label{sec:intro}
Small cells is considered the most promising technique to keep up with the exponential increase of data-rate demand foreseen for 5G networks \cite{5GJSAC}. However, more base stations (BSs) sharing the same spectrum result in increased multi-cell interference, which is a major limiting factor if not properly managed \cite{Gesbert2010}.
Cooperative Multi-Point (CoMP), already standardized in long term evolution advanced (LTE-A) \cite{CoMPLTE}, aims at turning inter-cell interference (ICI) into an advantage by letting BSs share their data and perform joint precoding/decoding. This requires the exchange of global channel state information (CSI) as well as (possibly) user data via high data rate backbone connections, which might be a problem when the BSs belong to different operators or have conflicting utilities. In these cases, coordination schemes among BSs without global CSI and user data exchange might be feasible \cite{CoMPMagzine2011}.

In this work, we consider an interfering multiple-input and multiple-output (MIMO) multiple access channel (MAC) network, which is well matched to the multi-cell multi-user uplink scenario. Multiple cells share their spectrum so as to form a coordinated cluster. Each BS serves multiple users within its own cell and each node is equipped with multiple antennas. The uplink signal is corrupted by both ICI and inter-user interference (IUI). In order to eliminate both interference terms, simple \emph{linear transceiver} implementations are preferred. This was addressed in \cite{Kim2010} by applying a coordinated zero-forcing (ZF) scheme to mitigate both IUI and ICI in the interfering MIMO broadcast channel (BC). However,  ZF alone fails if a BS does not have sufficient antennas or if degrees of freedom (DoF) maximization is the goal.
With this respect, a well-established technique called interference alignment (IA) is helpful \cite{NowIA, IACaJafar08}.
IA is applied to suppress the interference at a given receiver, thereby reducing the number of antennas required to implement ZF reception \cite{Choi2009}.
IA for DoF and sum-rate optimization in $K$-user MIMO interference channel is considered in \cite{Sung2010, SumRateIA10, IAGomadamJafar11} by designing the linear IA precoders and decoders. 
Generally, it is difficult to obtain the closed-form linear IA transceiver and iterative algorithms based on global CSI are usually required, except for the special case of square and invertible channel matrices, e.g. in  \cite{Sung2010}.
More recently, IA has been applied to MIMO cellular networks. In \cite{nagarajan2010}, a multi-cell MIMO downlink channel is studied and a distributed IA algorithm is proposed to suppress or minimize the interference to non-intended users. Also, \cite{Suh2011} develops an IA technique for a downlink cellular system with CSI-exchange and feedback within each cell. In \cite{IAYetisJafar10, DoFMIMOMAC2011}, conditions for the feasibility of IA and DoF for MIMO cellular networks are investigated. To reduce the complexity and CSI requirement, the concept of \emph{grouping-based} IA (GIA) is proposed for a two-cell single-stream interfering MIMO-BC in \cite{Shin2011}. The idea is to let each cell align its interference to another cell, which will then require less antennas to implement ZF reception. Moreover, the GIA enables to compute the closed-form IA transceiver based on only local CSI.
 This GIA is extended to a multi-cell interfering MIMO-BC in \cite{Tang2013}, where both the feasible condition on the GIA and a low complexity IA decoder design are studied. In addition, some works extend the GIA to the limited feedback scenario in the two-cell single-stream interfering MIMO-MAC, e.g., \cite{GIAMIMOMAC1,GIAMIMOMAC2,GIAMIMOMAClimited3}.

The implementation of IA requires a closed-loop transmission. The feedback is needed in either the downlink or the uplink scenario\footnote{In the downlink, the feedback takes two phases: 1) the downlink CSI is first fed back to BSs and 2) the IA decoders designed at BSs are reported to users (also called dedicated training phase). In the uplink, the IA precoders designed at BSs based on the perfect CSIR are fed back to the users.}. 
Since the feedback links are usually capacity-limited in realistic scenarios, codebook-based feedback is widely used and already defined in modern wireless standards, e.g., in LTE \cite{CodebookbasedLTE}, to reduce the feedback overhead. The idea is to map a channel matrix/vector or precoder/decoder to an index of the closest codeword in a predefined codebook known at both transmitter and receiver. The feedback of an index takes only a limited number of feedback bits, while a performance loss is inevitable because of the quantization distortion. Thus, it becomes an important issue how to control/reduce the performance loss under limited feedback \cite{Love2008}. 
For a MIMO BC with ZF precoder, the performance loss due to limited feedback is studied in \cite{Jindal2006, Yoo2007} and also with block diagonalization in \cite{Ravindran2008, Schwarz2013}. 
For a MIMO interference channel with heterogeneous path loss and spatial correlations, \cite{Rao2013} develops a spatial codebook design and performs a subspace quantization scheme via feedback bit allocation. In \cite{Bolsckei2009, RIAKrishnamachari13, Rezaee2012}, the feedback bits scaling law to maintain the maximum DoF for IA on general MIMO interference networks is investigated. 

Motivated by this background, we focus on the GIA in a multi-cell interfering MIMO-MAC under limited feedback, answering the following fundamental questions:
\begin{enumerate}
\item \textbf{How to design the optimal linear GIA transceiver with low complexity?}  
We further develop previous related works (e.g. \cite{Shin2011, Tang2013}), providing a low-complexity \emph{restriction-and-relaxation} approach to compute the linear GIA transceivers which not only nulls out both ICI and IUI but also maximizes the rate performance. The tightness of the proposed restriction-and-relaxation procedure is verified, which implies that the computed IA transceiver is optimal.

\item \textbf{How to determine a good IA-Cell assignment?}
By the GIA, each cell chooses to align its interference to another cell. However, this choice clearly impacts the rate performance. Optimizing the selection of the cell to/from which a given cell provides/receives the aligned interference, is a problem which was not considered in previous works. We refer to this problem as \emph{IA-Cell assignment} and provide three IA-Cell assignment algorithms: a centralized one, which yields global optimality but requires high complexity and backhaul overhead, and two distributed ones, which yield a stable or almost stable assignment with limited complexity and backhaul overhead. 

\item \textbf{How to efficiently feed back the GIA precoders to the transmitters?} In the uplink MIMO cellular scenario, the GIA precoders need to be fed back to the users. 
We employ Grassmannian subspace quantization, developing a novel quantized subspace characterization which allows one to derive a closed-form upper bound of the single-cell residual interference to noise ratio (RINR). Based on this upper bound, we formulate and solve in closed-form a feedback bit allocation problem for sum-cluster RINR minimization. Furthermore, the effect of the sum feedback bit budget on the sum-cluster rate is analyzed.
\end{enumerate} 

The three contributions above jointly provide a comprehensive holistic design of the multi-cell MIMO MAC system under limited feedback. 

The paper is organized as follows: a complete study of the GIA on DoF and optimal linear transceiver design is provided in Section \ref{sec:ICA}. In Section \ref{sec:Matching}, the IA-Cell assignment problem is addressed and solved. The limited feedback scenario is considered in Section \ref{sec:feedback}. In Section \ref{sec:analysis}, we analyze the practical implementation, backhaul overhead requirements and complexity of the proposed GIA algorithm with optimized IA-Cell assignment and under limited feedback. The numerical results in Section \ref{sec:simulations} show the effectiveness of the proposed algorithms under unlimited and limited feedback.\footnote{\emph{Notations:} 
$\mathbb{N}_0^{+}$ denotes the nonnegative integer domain. $e$ denotes the Euler's number.  $[x]_{int}$ and $[x]_{int}^+$ denote the integer and the nonnegative integer around $x$, respectively. Give a $M\times 1$ vector $\mat{x}$, $\mathrm{arglist}~\max_{m=1, \ldots, M}~ \mat{x}$ generates a $M\times 1$ vector where the elements are re-arranged in decreasing order.
$(\cdot)^H$, $\mathrm{rank}(\cdot)$ and $\mathrm{Tr}(\cdot)$ denote Hermitian transpose, rank and trace, respectively.
$\lambda_i(\mat{X})$ and $\mat{U}_{{\mat{X}}}$ denote the $i$-th \emph{largest} eigenvalue and the eigen-space of $\mat{X}$, respectively.
$\mathrm{Span}\{\mat{X}\}$ denotes the space spanned by the column space of $\mat{X}$. 
$\Pi_{\mat{X}}\stackrel{\Delta}{=}{\mat{X}}({\mat{X}^H}{\mat{X}})^{-1}{\mat{X}}^H$ denotes the orthogonal projection onto the column space of ${\mat{X}}$,
and $\Pi_{\mat{X}}^{\perp}\stackrel{\Delta}{=}\mat{I} - \Pi_{\mat{X}}$ denotes the orthogonal projection onto the orthogonal complement of the column space of ${\mat{X}}$.
$\mat{X}^{\bot}$ is defined as the \emph{left null space} of the matrix $\mat{X}$, i.e., the eigen-subspace spanned by the eigenvectors associated with those zero-eigenvalues of $\mat{X}\mat{X}^H$, such that $(\mat{X}^{\bot})^H\mat{X} = \mat{0}$. 
}

\section{System Model} \label{se: SM}

Consider a MIMO cellular environment with $K$ cells.
In each cell, a central BS simultaneously serves $L$ users in its own cell, where each BS and each user are equipped with $N_B$ and $N_U$ antennas, respectively.
In order to increase the spectral efficiency and occupancy level compared with classical FDMA and TDMA techniques, $K$ cells form a \emph{coordinated cluster} and operate over the same time-frequency resource, while the introduced IUI and ICI in return corrupt the received desired signal and limit the detection efficiency or transmission rate. Thus, interference management is required.

This work focuses on the uplink scenario, where the setup is modeled as an interfering MIMO-MAC system $(K, L, N_B, N_U, d_s)$. Each user $i$ in cell $k$, denoted by user $(i,k)$, transmits $d_s$ symbols $\mat{x}_{i,k} \in \mathbb{C}^{d_s\times 1}$ with $\mathbb{E}[\mat{x}_{i,k}\mat{x}_{i,k}^H] = \mat{I}_{d_s} $ to its corresponding BS $k$. 
The symbol vector $\mat{x}_{i,k}$ is precoded by a linear precoder $\mat{V}_{i,k} \in \mathbb{C}^{N_U \times d_s}$ subject to $\trace(\mat{V}_{i,k}^H\mat{V}_{i,k})\leq P_{i,k}$ where $P_{i,k}$ is the transmit power budget.

We assume that the local CSIR is perfectly estimated at each BS based on the orthogonal uplink pilot signals. The received signal at BS $k$ for user $(i,k)$ is expressed as
\begin{align}
\mat{y}_{i,k} &= \underbrace{\mat{H}_{i,k}^k \mat{V}_{i,k} \mat{x}_{i,k}}_{\text{desired signal}} + \underbrace{\sum_{j=1, j\neq i}^{L}{\mat{H}_{j,k}^k \mat{V}_{j,k} \mat{x}_{j,k}}}_{\text{IUI}}  \nonumber \\
&+ \underbrace{\sum_{\ell=1, \ell \neq k}^K \sum_{m=1}^{L}{\mat{H}_{m,\ell}^k \mat{V}_{m,\ell} \mat{x}_{m,\ell}}}_{\text{ICI}} + \mat{n}_k, \label{eq:datayk}
\end{align}
where $\mat{H}_{i,k}^\ell$ denotes the channel matrix from user $(i,k)$ to BS $\ell$ and is modeled as $\sqrt{\eta_{i,k}^\ell }\overline{\mat{H}}_{i,k}^\ell$, where $\eta_{i,k}^\ell$ denotes the effect of path-loss, and $\overline{\mat{H}}_{i,k}^\ell \in \mathbb{C}^{N_B \times N_U}$ is a Rayleigh fading channel matrix.
 Each channel is assumed to be quasi-static and frequency flat fading.
$\mat{n}_k \in \mathbb{C}^{N_B \times 1}$ is the additive white Gaussian noise vector with zero mean and variance $\sigma_k^2 \mat{I}_{N_B}$.

With the linear single-user decoding scheme, the received signal vector $\mat{y}_{i,k}$ for user $(i,k)$ can be decoded as $\widehat{\mat{x}}_{i,k} = \mat{U}_{i,k}^H\mat{y}_{i,k}$ by the decoder $\mat{U}_{i,k} \in \mathbb{C}^{N_B \times d_s}$.
In order to make efficient detection of the desired signal, the desired signal should be linearly independent of the interference, i.e., 
the following conditions need to be satisfied:
\begin{subequations}
\begin{align}
\mat{U}_{i,k}^H\mat{H}_{j,k}^k \mat{V}_{j,k} &= \mat{0},~~~\forall j\neq i  \label{eq:IA1} \\
\mat{U}_{i,k}^H\mat{H}_{m,\ell}^k \mat{V}_{m,\ell} &= \mat{0}, ~~~\forall \ell \neq k,~\forall m   \label{eq:IA2} \\
\rank(\mat{U}_{i,k}^H\mat{H}_{i,k}^k \mat{V}_{i,k}) &= d_s,~~~\forall i,k, \label{eq:IA3}
\end{align}
\end{subequations}
where (\ref{eq:IA1}) and (\ref{eq:IA2}) enable the mitigation of IUI and ICI, respectively, and (\ref{eq:IA3}) guarantees the transmission of $d_s$ data streams per user.  
Then, the achievable rate for user $(i,k)$ is
\begin{align}
R_{i,k} =  \log_2\det\left(\mat{I}_{d_s} +  \frac{1}{\sigma_k^2} \mat{U}_{i,k}^H\mat{H}_{i,k}^k\mat{V}_{i,k}\mat{V}_{i,k}^H\mat{H}_{i,k}^{k,H}\mat{U}_{i,k}\right).  \label{eq:RateIA1}
\end{align}
For the conditions (\ref{eq:IA1})-(\ref{eq:IA3}) to be fulfilled in the system $(K, L, N_B, N_U, d_s)$, any user $(i,k)$ needs to satisfy
\begin{align}
&\mat{U}_{i,k}^H \Big[\{\mat{H}_{j,k}^k \mat{V}_{j,k}\}_{j=1, j\neq i}^L,~ {\{\mat{F}_{\ell}^k\}_{\ell=1, \ell \neq k}^K} \Big]  \nonumber \\
\triangleq ~&\mat{U}_{i,k}^H\mat{F}_{i,k} = \mat{0} \label{eq:IFL1}
\end{align}
where $\mat{F}_{i,k} \in \mathbb{C}^{N_B \times (KL-1)d_s}$ denotes the interference matrix. 

\emph{Sufficient and Necessary Conditions:} (\ref{eq:IFL1}) is fulfilled if and only if $N_B \geq \rank(\mat{F}_{i,k}) + d_s$ such that BS $k$ could provide at least a $\rank(\mat{F}_{i,k})$-dimensional subspace to nullify all the interference to user $(i,k)$ and simultaneously guarantee  $d_s$ DoF per user.  

Due to $\rank(\mat{F}_{i,k}) \leq (KL-1)d_s$,  it is sufficient to fulfill (\ref{eq:IFL1})  by only exploiting the ZF decoding if $N_B \geq KLd_s$. In general, we have $\rank(\mat{F}_{i,k}) = (KL-1)d_s$ if no restrictions is on the transmission through Rayleigh fading channels. 
In this paper, we study the interference mitigation in the non-trivial case $((K-1)L+1)d_s \leq N_B < KLd_s$ where the sole ZF decoding fails and IA is required.
Instead of developing iterative IA algorithms,
we deal with the problem of \emph{low-complexity} IA transceiver design, also considering the problem of IA-Cell assignment and limited feedback. 

For future reference, we first give the following definitions: 
The channel set from users in cell $k$ to BS $\ell$: $\mat{H}_k^{\ell} \triangleq \{\mat{H}_{i,k}^{\ell}\}_{i=1}^L$. The local CSIR of BS $\ell$: $\mat{H}^{\ell} \triangleq \{ \mat{H}_k^{\ell} \}_{k=1}^K$. 
The interference from cell $k$ to cell $\ell$: $\mat{F}_{k}^{\ell} \triangleq [\mat{H}_{1,k}^{\ell} \mat{V}_{1,k}, \ldots, \mat{H}_{L,k}^{\ell} \mat{V}_{L,k}]  \in \mathbb{C}^{N_B \times L d_s}$.
The IUI of user $(i, k)$: $\mat{F}_{i,k}^{IUI} \triangleq [\{\mat{H}_{j,k}^{k}\mat{V}_{j,k}\}_{j=1, j\neq i}^L]  \in \mathbb{C}^{N_B \times (L-1) d_s}$.

\section{Interference Alignment and Mitigation} \label{sec:ICA}

In this section, we develop a \emph{restriction-relaxation} two-stage algorithm based on the GIA method in \cite{Shin2011, Tang2013}, which enables to compute the optimal IA transceiver in closed-form.

\subsection{Feasible Conditions for the GIA}
The GIA method in \cite{Tang2013} is a generalization of the non-iterative grouping scheme originally proposed in \cite{Shin2011} to completely suppress the interference. 
\emph{The basic idea of the GIA method in \cite{Tang2013} is to group all the users in one cell to generate a joint precoder aligning their interference to another cell.} Let ${Cell~k} \stackrel{IA}{\longrightarrow} {Cell~k'}$ denote that cell $k$ aligns its interference to cell $k'$.  The feasible conditions for the GIA method and its DoF performance are shown in the following proposition.

\begin{proposition}\label{pr:DoF}
For a multi-cell interfering MIMO-MAC system $(K,L,N_B,N_U,d_s)$, at least $d_s$ DoF per user and $KLd_s$ sum DoF can be achieved by the GIA method if 
\begin{align}\label{eq:GIAcond}
N_U \geq \frac{L-1}{L}N_B + \frac{1}{L}d_s~ \mathrm{and}~ N_B \geq ((K-1)L +1)d_s. 
\end{align} 
\end{proposition}

\begin{IEEEproof}
Without loss of generality, to fix ideas we consider the following scenario. 
\begin{align}\label{eq:cellIASeassign}
{Cell~1} \stackrel{IA}{\longrightarrow} {Cell~2} \stackrel{IA}{\longrightarrow} \ldots \stackrel{IA}{\longrightarrow} {Cell~K}\stackrel{IA}{\longrightarrow} {Cell~1}.
\end{align}

In particular, the procedure of ${Cell~k} \stackrel{IA}{\longrightarrow} {Cell~k+1}$ can be implemented by finding $\{\mat{V}_{i,k}\}$ such that the following IA condition
\begin{align}
\overline{\mat{F}}_{k}^{k+1} \triangleq \mathrm{Span}\{\mat{H}_{1,k}^{k+1}\mat{V}_{1,k}\} = \ldots =\mathrm{Span}\{\mat{H}_{L,k}^{k+1}\mat{V}_{L,k}\}, \label{eq:IACh}
\end{align}
is fulfilled subject to the per-user transmit power constraint $\trace(\mat{V}_{i,k}\mat{V}_{i,k}^H) \leq P_{i,k}, \forall i=1,\ldots, L$.

Since the transmit power constraints do not influence the IA condition (i.e., subspaces alignment in \eqref{eq:IACh}), we first \emph{restrict} the IA condition (\ref{eq:IACh}) to find those precoding matrices such that
\begin{align}
\mat{H}_{1,k}^{k+1}{\mat{V}}_{1,k}^{in} =  \ldots =\mat{H}_{L,k}^{k+1}{\mat{V}}_{L,k}^{in}. \label{eq:IACh2}
\end{align}
In this \emph{restriction stage} (in fact, only on the "power" of $\mat{V}_{i,k}^{in}$),  (\ref{eq:IACh2}) is rewritten as
\begin{align}
&\begin{bmatrix}
\mat{H}_{1,k}^{k+1}  ~&-\mat{H}_{2,k}^{k+1}    ~&\mat{0}                      ~&\cdots    ~&\mat{0} \\
\vdots                     ~&\vdots                      ~&\vdots                       ~&\ddots     ~&\vdots  \\
\mat{H}_{1,k}^{k+1}  ~&\mat{0}                     ~&\mat{0}                     ~&\cdots      ~&-\mat{H}_{L,k}^{k+1}
\end{bmatrix}
\begin{bmatrix}
{\mat{V}}_{1,k}^{in} \\
{\mat{V}}_{2,k}^{in} \\
\vdots         \\
{\mat{V}}_{L,k}^{in}
\end{bmatrix}
\nonumber \\
&\triangleq \mat{A}_{k}^{k+1} {\mat{V}}_{k}^{in} = \mat{0} \label{eq:IACon3}
\end{align}
where $\mat{A}_{k}^{k+1} \in \mathbb{C}^{(L-1)N_B \times LN_U}$ and ${\mat{V}}_{k}^{in} \in \mathbb{C}^{LN_U \times d_s}$. To fulfill (\ref{eq:IACon3}), the joint IA precoder ${\mat{V}}_{k}^{in}$ should lie in the null space of $\mat{A}_{k}^{k+1}$, which requires $LN_U \geq (L-1)N_B + d_s$ such that $\mat{A}_{k}^{k+1}$ has a at least $d_s$-dimensional null space. 

By (\ref{eq:IACon3}), the original $Ld_s$-dimensional interference subspace of $\mat{F}_{k}^{k+1}$ is aligned to a $d_s$-dimensional subspace of $\overline{\mat{F}}_{k}^{k+1}$ because (\ref{eq:IACh}) holds, while the interference $\mat{F}_{k}^{\ell}~\forall \ell \neq k, k+1$ is still with $Ld_s$ dimensions.  For the scenario (\ref{eq:cellIASeassign}), it is sufficient for each BS $k$ to remove the complete interference for user $(i,k)$ by the ZF decoding if $N_B \geq ((K-1)L+1)d_s$. 
\end{IEEEproof}

\begin{remark}
By the feasible conditions (\ref{eq:GIAcond}) in Proposition \ref{pr:DoF}, we gain the following insights on system design.
\begin{itemize}
     \item[{1})] Given $(K,L,N_B,N_U)$, each user achieves at most $\min(LN_U - (L-1)N_B, \frac{N_B}{(K-1)L+1})$ DoF;
      \item[{2})] Given $(K,L,N_B,d_s)$, each user needs at least $((L-1)(K-1)+1)d_s$ antennas to guarantee its $d_s$ DoF;
      \item[{3})] Given $(K,N_B,N_U,d_s)$, each cell serves at most $\min(\frac{N_B-d_s}{N_B-N_U}, \frac{N_B-d_s}{(K-1)d_s})$ users;
      \item[{4})] Given $(L,N_B,N_U,d_s)$, at most $\frac{N_B-d_s}{Ld_s} +1$ cells can be scheduled to form a cluster with the sum DoF of $KLd_s$ if $N_U\geq \frac{L-1}{L}N_B + \frac{1}{L}d_s$.
   \end{itemize}

If the inequalities in both feasible conditions (\ref{eq:GIAcond}) become equalities, the required number of BS and user antennas are the smallest.
\end{remark}

\subsection{Transceiver Optimization for the GIA}\label{subset:transceiver}
As in \cite{Tang2013,Cupta2013}, we hereafter focus on the worst-case that $N_B = ((K-1)L+1)d_s$ and $N_U = \lceil\frac{L-1}{L}N_B + \frac{1}{L}d_s\rceil$.  
In this case, the optimal GIA transceiver are computed in closed-form.

\begin{proposition}\label{pr:LTD}
Let us define
\begin{align}
\mat{T}_i &\triangleq [\mat{0}_{N_U\times (i-1)N_U}, \mat{I}_{N_U}, \mat{0}_{N_U\times (L-i)N_U}]  \label{eq:Tdef}  \\
\mat{V}_k^{in} &= \left(\mat{A}_{k}^{k+1, H}\right)^{\perp}    \label{eq:Vkindef}  \\
\mat{F}_{i,k}^{IA,k-1} &\triangleq \left[{\mat{F}_{i,k}^{IUI} ,~{\left\{\mat{F}_{\ell}^{k}\right\}_{\ell=1, \ell \neq k, k-1}^K}},~{\overline{\mat{F}}_{k-1}^{k}} \right]. \label{eq: FIAdef}
\end{align}
Considering (\ref{eq:cellIASeassign}) and the uniform power allocation policy, the achievable rate of each user $(i,k)$ in (\ref{eq:RateIA1}) is maximized by 
the optimal transceiver
\begin{align}
&\mat{V}_{i,k} = \sqrt{\frac{P_{i,k}}{d_s}} \mat{T}_{i}{\mat{V}}_{k}^{in}({\mat{V}}_{k}^{in, H}\mat{T}_{i}^H\mat{T}_{i}{\mat{V}}_{k}^{in})^{-\frac{1}{2}}   \label{eq:Vikoptimal} \\ 
&\mat{U}_{i,k} = \left(\mat{F}_{i,k}^{IA,k-1}\right)^{\perp}. \label{eq:Uikoptimal}
\end{align}
\end{proposition}   
\begin{IEEEproof}
Without loss of generality, we consider the scenario (\ref{eq:cellIASeassign}). First, we observe that $\mat{V}_{k}^{in}$ must lie in the null space of $\mat{A}_{k}^{k+1}$ to fulfill (\ref{eq:IACon3}), thereby \eqref{eq:Vkindef}. Based on the fact $\mathrm{Span}({\mat{V}}_{i,k}^{in} \mat{X}) = \mathrm{Span}({\mat{V}}_{i,k}^{in})$ where $\mat{X}\in\mathbb{C}^{d_s \times d_s}$ is an arbitrary full-rank matrix variable, the IA precoder for each user $(i,k)$ can be defined as
\begin{align}
\mat{V}_{i,k} \triangleq {\mat{V}}_{i,k}^{in} \mat{V}_{i,k}^{out} =\mat{T}_{i}{\mat{V}}_{k}^{in} \mat{V}_{i,k}^{out}   \label{eq:Vik2}
\end{align}
where $\mat{T}_i$ is a selection matrix defined in (\ref{eq:Tdef})  and $\mat{V}_k^{in}$ is an \emph{inner precoder} defined in (\ref{eq:Vkindef}), and $\mat{V}_{i,k}^{out}\in\mathbb{C}^{d_s \times d_s}$ is an \emph{outer precoder} subject to the transmit power constraint $\trace(\mat{V}_{i,k}^{out,H}{\mat{V}}_{i,k}^{in,H}{\mat{V}}_{i,k}^{in}\mat{V}_{i,k}^{out}) \leq P_{i,k}$, which is used to \emph{relax} the "power" restriction from (\ref{eq:IACh}) to (\ref{eq:IACh2}), since the transmit power constraint was not jointly considered in the restriction stage. 
The optimal precoder $\mat{V}_{i,k}$ can be determined by further optimizing ${\mat{V}}_{i,k}^{out}$. 

Also due to $\mathrm{Span}(\mat{H}_{j,\ell}^{k}{\mat{V}}_{j,\ell}^{in}\mat{V}_{j,\ell}^{out}) = \mathrm{Span}(\mat{H}_{j,\ell}^{k}{\mat{V}}_{j,\ell}^{in})$, it is sufficient to design the ZF decoder $\mat{U}_{i,k}$ only based on ${\mat{V}}_{j,\ell}^{in}$ but without knowledge of $\mat{V}_{j,\ell}^{out}$. 
The ZF decoder for user $(i,k)$ can be designed to nullify the total received interference by
\begin{align}
\mat{U}_{i,k} = \left(\mat{F}_{i,k}^{IA,k-1}\right)^{\perp},  \label{eq:Uik0}
\end{align}
where $\mat{F}_{i,k}^{IA,k-1}$ defined in (\ref{eq: FIAdef}) is a $N_B \times (K-1)Ld_s$ interference matrix with the aligned interference from cell $k-1$.

With the IA transceiver in form of (\ref{eq:Vik2}) and (\ref{eq:Uik0}), the achievable rate of each user $(i, k)$ becomes
\begin{align}
R_{i,k}^{IA} = \log_2\det\Big( \mat{I}_{d_s} + \frac{1}{\sigma_{k}^2}\widetilde{\mat{H}}_{i,k}^k\mat{V}_{i,k}^{out}\mat{V}_{i,k}^{out, H} \widetilde{\mat{H}}_{i,k}^{k,H}\Big), \label{eq:RIAnew}
\end{align}
where $\widetilde{\mat{H}}_{i,k}^k$ is the effective channel from user $(i,k)$ to BS $k$
\begin{align}
\widetilde{\mat{H}}_{i,k}^k \triangleq \mat{U}_{i,k}^{H}\mat{H}_{i,k}^k\mat{T}_{i}\mat{V}_{k}^{in}.
\label{eq:EffChannel}
\end{align}
We observe that after perfect interference mitigation, $\{\mat{V}_{i,k}^{out}\}$ are decoupled across the users as shown in \eqref{eq:RIAnew}. 
Then, $\{\mat{V}_{i,k}^{out}\}$ can be optimally computed by maximizing the individual rate $R_{i,k}^{IA}$ in \eqref{eq:RIAnew} subject to the power constraints $\trace(\mat{V}_{i,k}^{out,H}{\mat{V}}_{i,k}^{in,H}{\mat{V}}_{i,k}^{in}\mat{V}_{i,k}^{out}) \leq P_{i,k}$, where $\mat{V}_{i,k}^{in}$ is given in \eqref{eq:Vkindef}. Clearly, the optimal $\mat{V}_{i,k}^{out}$ should diagonalize $\widetilde{\mat{H}}_{i,k}^{k,H}\widetilde{\mat{H}}_{i,k}^{k}$ maybe with the standard water-filling power allocation (if $\rank(\mat{V}_{i,k}^{out} )= d_s$ to support the $d_s$ data streams per user). 
Due to practical considerations, we assume uniform power allocation{\footnote{Instead of the water-filling based power allocation across the data streams, the uniform power allocation policy is adopted because of the following reasons: {1}) it is known to be asymptotically optimal for large SNR \cite{Ravindran2008}, {2}) it guarantees the transmission of $d_s$ data streams per user (i.e., condition (\ref{eq:IA3})),  {3}) it has lower complexity compared with water-filling process and {4}) it is not necessary to feed back the outer precoders to users.}}. It yields ${\mat{V}}_{i,k}^{out}=\sqrt{\frac{P_{i,k}}{d_s}}(\mat{V}_{i,k}^{in,H}\mat{V}_{i,k}^{in})^{-\frac{1}{2}}$, thereby (\ref{eq:Vikoptimal})-(\ref{eq:Uikoptimal}).
\end{IEEEproof}

The improvements of the derived results with respect to previous works on the GIA  \cite{Shin2011, Tang2013} are two-fold.
\begin{itemize}
\item \emph{Lower complexity:} The complexity of the GIA mainly depends on the singular-value decomposition (SVD) of $K$ matrices $\{\mat{A}_{k}^{k+1}\}$. By the \emph{new formulation} (\ref{eq:IACon3}), our GIA takes $K\mathcal{O}((L-1)^2LN_B^2 N_U)$ arithmetic operations, since each $\mat{A}_{k}^{k+1}$ is a $(L-1)N_B \times L N_U$ matrix. 
In contrast, \cite[Eq. (27)]{Tang2013} (same as \cite{Shin2011}) and \cite[Eq. (12)-(13), (15)]{Tang2013} have the complexity of $K\mathcal{O}(L^2N_B^2(LN_U + N_B))$ or $K\left(L\mathcal{O}(N_B^3 + N_B^2N_U) + 2(L+\log_2(L))\mathcal{O}(2N_B^2N_U)\right)$, respectively. It follows that the complexity of our GIA by (\ref{eq:IACon3}) is always lower than \cite[Eq. (27)]{Tang2013} and also lower than that by \cite[Eq. (12)-(13), (15)]{Tang2013} when $L \leq 3$.\footnote{The computation of the left singular-space and the singular values of a $M\times N$ matrix where $M<N$ is $4NM^2 + 8M^3$ arithmetic operations \cite{ComplexitySVD}. Based on this the complexity comparison with \cite{Tang2013} is done.}

\item \emph{Tightness of the restriction-and-relaxation:} In this work, we design $\{\mat{V}_{i,k}^{in}\}$ and $\{\mat{V}_{i,k}^{out}\}$, respectively, in two stages: 
1) design $\mat{V}_{i,k}^{in}$ based on the  IA condition restricted from (\ref{eq:IACh}) to (\ref{eq:IACh2}) (the restriction stage) and 2) design $\mat{V}_{i,k}^{out}$ subject to the power constraint $\trace(\mat{V}_{i,k}^{out,H}{\mat{V}}_{i,k}^{in,H}{\mat{V}}_{i,k}^{in}\mat{V}_{i,k}^{out}) \leq P_{i,k}$ (the relaxation stage). Such a procedure is termed here by the \emph{restriction-relaxation two-stage procedure}.
The tightness of the proposed restriction-and-relaxation procedure is proved by the following lemma.
\begin{lemma}\label{le:tight}
The proposed restriction-and-relaxation two-stage procedure -- design $\{\mat{V}_{i,k}^{in}\}$ and $\{\mat{V}_{i,k}^{out}\}$ separately in two stages -- is tight in GIA precoder design.
\end{lemma}
\begin{IEEEproof}
Please refer to the proof in Appendix \ref{subsec:Prooftight}.
\end{IEEEproof}
This tightness guarantees the optimality of the GIA-based linear transceiver in Proposition \ref{pr:LTD} under the uniform power allocation policy, which is designed based on the restriction-relaxation scheme.
\end{itemize}

\begin{remark}
The GIA as a non-iterative algorithm determines the IA transceiver in a distributed way and with low complexity.
For the distributed implementation, BSs need to exchange their inner precoders $\{\mat{V}_k^{in}\}_{k=1}^K$ with each other, while
the outer precoder $\mat{V}_k^{out}$ can be designed by each user $(i,k)$ independently. 
\end{remark}

\section{IA-Cell Assignment Problem Formulations and Solutions} \label{sec:Matching}
In this section we introduce the concept of IA-Cell assignment, motivate its importance for network performance and propose three algorithms for assignment optimization.

\subsection{IA-Cell Assignment Problems}

\subsubsection{Observation and Motivation}
For ${Cell~k} \stackrel{IA}{\longrightarrow} {Cell~k'}$, we label cell $k$ as the \emph{IA-provider} for cell $k'$ and cell $k'$ as the \emph{IA-receiver} from cell $k$. Clearly, this poses an assignment problem between IA-providers and IA-receivers -- how should we select the IA-receiver (or IA-provider) corresponding to a given IA-provider (or IA-receiver)?
From the perspective of spatial resources, a cell will waste part of its transmit spatial resources if it aligns its interference to other cells because of the IA constraint. On the other hand, a cell can save its receive spatial resource if it receives the aligned interference from other cells. 
Thus, providing IA and receiving IA can be considered as \emph{costs} and \emph{gains}, respectively. 
In order to gain mutual benefits, it is expected that each cell in a coordinated cluster simultaneously serves as an IA-provider and IA-receiver (i.e., \emph{gains with costs}). This is motivated by fairness reasons and allows for distributed implementations and self-organization.
The mapping of $K$ potential aligned interference to $K$ cells in a coordinated cluster can be formulated as an \emph{IA-Cell assignment} problem. 
 Now, two questions arise: 
 \begin{itemize}
 \item[Q1:] \emph{How many possible IA-Cell assignments exist in a $K$-cell cluster?} 
 \item[Q2:] \emph{How to find a good IA-Cell assignment?} 
\end{itemize}

\subsubsection{Effect of Assignment on DoF}
In order to answer the above questions, we start with the following definitions regarding the IA-Cell assignment.
\begin{definition}(Coordinated Cell and Lone Cell)
If a cell receives the aligned interference from other cells and it also aligns its own interference to others, this cell is called a \emph{coordinated cell}; Otherwise, a cell is called a \emph{lone cell} if it does not receive an IA from others and also it has no incentive to and will not provide its IA to others. 
\end{definition}

\begin{definition}(Strict/Weak IA-Cell Assignment)
The assignment is called a \emph{strict IA-Cell assignment} if each cell is a coordinated cell, e.g., the example in (\ref{eq:cellIASeassign}). 
Otherwise, we have a \emph{weak IA-Cell assignment}.
\end{definition}

For the considered system $(K, L, N_U, N_B)$, maximum DoF can be achieved only under the strict IA-Cell assignment, which can be easily proved by contradiction. 
Otherwise, the lone cell has to reduce its transmit data streams because it receives $(K-1)Ld_s$-dimensional interference and thus its desired $Ld_s$ DoF cannot be supported by $N_B=(K-1)Ld_s + d_s$ receive antennas. Under a weak IA-Cell assignment, the lone cell has only $d_s$ DoF, while other coordinated cells are with $Ld_s$ DoF per cell. 
For instance, the system $(K, L, N_U, N_B)=(3, 2, 6, 10)$ can achieve 12 sum DoF (4 DoF per cell) under a strict IA-Cell assignment, while only $10$ sum DoF is achieved when there exists a lone cell ($4$ DoF per coordinated cell and $2$ DoF of the lone cell). Therefore, \emph{a lone cell is suboptimal} if either the sum DoF or fairness is concerned. 
Thus, the focus will be on the strict IA-Cell assignment from now on.  

Question Q1 is answered by the following lemma.

\begin{lemma}\label{le:derangement}
A $K$-cell IA-Cell assignment problem where $K\geq 3$ has $K!\sum_{k=0}^K\frac{(-1)^k}{k!}-1$ strict IA-Cell assignments  in total. 
\end{lemma}

\begin{IEEEproof}
Let us label $K$ cells with the index sequence $1, 2, \ldots, K$. 
Under a strict IA-Cell assignment,
each cell simultaneously serves as an IA-provider and IA-receiver and both for other cells.
Therefore, the index sequence of $K$ IA-providers or IA-receivers of 
the $K$ cells in the sequence of $1, 2, \ldots, K$ 
should not share the same index at a common position. It can be formulated as a well-known \emph{derangement problem: determine  the permutations of the $K$ elements of a set such that none of the elements appear in their original positions}, 
which has $K!\sum_{k=0}^K\frac{(-1)^k}{k!}$ derangements \cite{Hassani2003}.
\end{IEEEproof}

\begin{corollary} \label{co: assignmentDoFs}
Under different strict IA-Cell assignments, the system $(K, L, N_U, N_B)$ has the same DoF performance. 
\end{corollary}
\begin{IEEEproof}
Under an arbitrary strict IA-Cell assignment, the dimension of the space spanned by the interference to and from each BS is the same. 
Therefore, Corollary \ref{co: assignmentDoFs} is concluded in the homogeneous system. 
\end{IEEEproof}

\subsubsection{Effect of Assignment on Rate Performance}
Different strict assignments have the same DoF, but they have different rate performance, because
the achievable rate (\ref{eq:RIAnew}) is determined by the effective channel $\widetilde{\mat{H}}_{i,k}^k\widetilde{\mat{H}}_{i,k}^{k,H}$.
This effective channel highly depends on the IA-Cell assignment, because $\mat{V}_{i,k}^{in}$ and $\mat{U}_{i,k}$ are thin matrices and could select multiple possible singular-values (or their combinations) of $\mat{H}_{i,k}^k$ in (\ref{eq:EffChannel}), and thus they are varying with the IA-Cell assignment. 

Inspired by (\ref{eq:EffChannel}), each cell should have double preferences:  the \emph{IA-provider preference}  and the \emph{IA-receiver preference}, based on which each cell could find its preferred IA-receiver and IA-provider. However, it is not possible to determine the optimal preferences before assignment because they are \emph{coupled}: 1) the preferences of one cell depend on other cells' assignment and 2) the IA-provider preference and IA-receiver preference of an individual cell depend on each other. Even if the approximate preferences are available, there is still a problem -- how to balance the conflicts of multiple cells when some of them have the same preferred objective.  
In order to make the problem solvable and answer question Q2, we consider three scenarios with different practical constraints (e.g., different backhaul overhead and coordination levels), and apply the stable matching or centralized assignment to obtain a stable or optimal strict IA-Cell assignment for each scenario. 

As a desired criterion, the stability of the IA-Cell assignment can be defined as follows.
\begin{definition}(Stable Assignment)
An IA-Cell assignment is \emph{stable} if there does not exist a subset of cells consisting of more than one cell, in which the reassignment of IAs makes at least one cell better off but none worse off than their current assignment.
\end{definition}

\subsection{One-Sided IA-Cell Matching}
In this part, we consider the case when \emph{no backhaul overhead is allowed between BSs before assignment.} In this case, each BS determines its assignment only based on its local CSIR.

\subsubsection{{Preference Generation}}
Since each BS $k$ only knows its desired channels $\mat{H}_{k}^k$ and interference channels $\{\mat{H}_{\ell}^k\}_{\ell \neq k}$, it can compute $K-1$ \emph{potential} IA precoders        
$\{\mat{V}_{\ell}^{in}(k)\}_{\ell \neq k}$ for the $K-1$ cells (\emph{potential IA-providers}) based on $\{\mat{H}_{\ell}^k\}_{\ell \neq k}$, where $\mat{V}_{\ell}^{in}(k), \forall \ell \neq k$ denotes the potential IA precoder for cell $\ell$ if cell $\ell$ serves as the IA-provider for cell $k$, which can be computed at BS $k$ by \eqref{eq:IACon3} and \eqref{eq:Vkindef} based on the CSI $\mat{H}_{\ell}^k$. Under a strict IA-Cell assignment, each BS has only one IA-provider, and thus each BS $k$ needs to rank the $K-1$ potential IA-providers by evaluating their corresponding interference subspace $\{{\overline{\mat{F}}}_{i,k}^{\ell}\}$, where ${\overline{\mat{F}}}_{i,k}^{\ell}$ is the complete interference subspace for user $(i, k)$ when cell $\ell$ is the IA-provider for cell $k$ defined in \eqref{eq: FIAdef}. However, each BS cannot construct the complete interference subspace because it does not know the IA precoders of all cells before assignment but only the potential IA precoders from its potential IA-providers.  
Therefore, BS $k$ cannot determine its IA-receiver preference but its IA-provider preference based on the $K-1$ potential aligned interference subspaces $\{{\overline{\mat{F}}}_{\ell}^{k}\}_{\ell \neq k}$, where $\overline{\mat{F}}_{\ell}^{k}, \forall \ell \neq k$ denotes the aligned interference from cell $\ell$ to cell $k$ as shown in \eqref{eq:IACh}. 

Let $\mathcal{P}_k^{p}$ with $K-1$ elements be arranged in decreasing order be the IA-provider preference list of BS $k$, i.e., 
\begin{align}
\mathcal{P}_k^{p}  
= \mathrm{arglist}~\max_{\ell\neq k}~\sum_{i=1}^L \log_2\det\left( \mat{I}_{N_U} + (\mat{H}_{i,k}^k)^H\mat{\Pi}_{\overline{\mat{F}}_{\ell}^k}^{\perp} \mat{H}_{i,k}^k \right).
\label{eq:Preflist1}
\end{align}
The performance metric{\footnote{The performance metric in (19) is derived from $\sum_{i=1}^L \log_2\det( \mat{I}_{N_U} + \mat{H}_{i,k}^{k,H}\mat{\Pi}_{\overline{\mat{F}}_{\ell}^k}^{\perp, H} \mat{\Pi}_{\overline{\mat{F}}_{\ell}^k}^{\perp}\mat{H}_{i,k}^k )$ based on the following properties $\mat{\Pi}_{\mat{X}}^{\perp, H} = \mat{\Pi}_{\mat{X}}^{\perp}$ and  $\mat{\Pi}_{\mat{X}}^{\perp, H} \mat{\Pi}_{\mat{X}}^{\perp} = \mat{\Pi}_{\mat{X}}^{\perp}$.}} in (\ref{eq:Preflist1}) is to approximately measure the effect of the potential aligned interference subspace on the sum rate of cell $k$ without knowledge of its own IA precoders. Note that by \eqref{eq:Preflist1} each BS has a single \emph{incomplete preference list}, which excludes itself because it does not desire to be a lone cell.

\subsubsection{{Modified Residence Exchange Model based IA-Cell Matching}}

The one-sided matching problem is modeled by the \emph{stable residence exchange model} \cite{Yuan1996} in which $K$ families wish to exchange their residences.
Each family has a move-in preference list consisting of up to $K$ choices \emph{with the least choice being its own residence without change}. The stable residence exchange demands that each family owns only one residence and each residence can only be rented by one family. This allocation involves a one-to-one matching between $K$ families and $K$ residences. Interpreting cells as families, IAs as residences, and IA-exchange as residence-exchange, our IA-Cell assignment will be well-matched to the stable residence exchange model if its \emph{incomplete preferences} can be relaxed by allowing the existence of a lone cell.

\emph{\textbf{a) Relaxation to Weak IA-Cell Assignment:}}
First, we relax our strict IA-Cell assignment to the \emph{weak IA-Cell assignment} by adding itself as the last candidate in the preference list of each BS. Then, 
the algorithm originally called the \emph{Top Trading Cycle Method} in \cite{Shapley1974} and renamed as the \emph{Forward Chaining Algorithm (FCA)} in \cite{Yuan1996} always generates a unique stable solution for this weak IA-Cell assignment problem.

For ${Cell~k'} \stackrel{IA}{\longrightarrow} Cell~k$, a cycle chain, denoted as $\left<Cell~k, Cell~k'\right>$, is formed if ${Cell~k} \stackrel{IA}{\longrightarrow} Cell~k'$. 
\emph{The basic idea of the FCA is to let each cell sequently choose its {current} most preferred until a cycle chain is formed.} By the FCA \cite{Yuan1996}, a stable weak IA-Cell assignment can be always obtained. 
\begin{corollary} \label{pr:FCA}
For a $K$-cell weak IA-Cell assignment, a stable solution always exists and is unique; The solution generated by the FCA is stable; No cell can be better off by misrepresenting its true preferences, assuming other cells keep their preferences unchanged. Even when several cells try to collude by misrepresenting their true preferences, it is impossible to make at least one better off and none worse off among themselves.
\end{corollary}
\begin{IEEEproof}
See \cite{Yuan1996,Shapley1974}.
\end{IEEEproof}

\begin{corollary}\label{co:FCASolu}
For a $K$-cell weak IA-Cell assignment, the stable matching by the FCA must fall in one of two cases: 1) no cell is lone cell; 2) only one cell is a lone cell.
\end{corollary}

\begin{IEEEproof}
This corollary can be easily proved by contradiction. Assume that there exist two lone cells. Since each cell has a complete IA-provider preference list where the cell itself is the last choice, these two lone cells surely prefer to exchange IA with each other rather than keep them. 
\end{IEEEproof}
\begin{remark}
If a stable matching for the weak IA-Cell assignment has no lone cell, this matching is also stable for the strict IA-Cell assignment. Otherwise, the strict IA-Cell assignment has no stable matching. 
\end{remark}

\emph{\textbf{b) "Almost Stable" Matching{\footnote{For the assignment problem, if a stable matching does not exist, it is desired to match as many pairs as possible, i.e., to find a matching with maximum cardinality (i.e., an as stable as possible matching)\cite{Biro2010}.}} by a Breaking Step:}}
When the stable weak IA-Cell assignment has a lone cell, 
the $K-1$ coordinated cells find their preferred IA-providers and each achieves $Ld_s$ DoF, but the lone cell with only $d_s$ DoF may reject to join the cluster because its desired $Ld_s$ DoF cannot be supported. This in return may degrade the $K-1$ coordinated cells' rate performance due to losing the spectrum or time resource shared by the lone cell. 
To circumvent this drawback, we modify the FCA by \emph{allowing the possibility to break a cycle and insert the lone cell to form a new larger cycle (breaking step)} such that each cell achieves $Ld_s$ DoF.
In this case, an "almost stable" matching always has a better DoF performance than the stable weak matching with a lone cell. Additionally, it may also improve the sum-utility performance, as shown in the following toy example. 
In Table \ref{table:AlmostStable1}, by the FCA, a stable weak IA-Cell assignment is first achieved, i.e., $<Cell~1, Cell~3, Cell~2>$ and $<Cell~4>$. Then, by the breaking step, we insert the lone cell $Cell~4$ into the cycle chain ${Cell~3} \stackrel{IA}{\longrightarrow} Cell~1\stackrel{IA}{\longrightarrow} Cell~2\stackrel{IA}{\longrightarrow} Cell~3$, e.g., by forcing the lone cell to choose its best preferred one, thereby forming an extended cycle $Cell~1\stackrel{IA}{\longrightarrow} {Cell~4} \stackrel{IA}{\longrightarrow} {Cell~3} \stackrel{IA}{\longrightarrow} Cell~2\stackrel{IA}{\longrightarrow} Cell~1$. This "almost stable" assignment with sum utility of $3+1+3+3=10$ and $4Ld_s$ sum DoF outperforms the original matching by the FCA only with the sum utility of $3+3+3+0=9$ and with $(3L+1)d_s$ DoF.

\begin{table}\caption{A toy example of $4$-cell assignment}
\centering
\begin{tabular}{l || l l l|l}
\hline
Cell      & \multicolumn{4}{l}{IA-Provider preference (utility)} \\ \cline{2-5}
         ~~~~&1st (3)    ~~~~&2nd (2)      ~~~~&3rd (1)    ~~~~&4th (0)  \\
\hline
1            &3   &2   &4  &1   \\
2            &1   &3   &4  &2   \\
3            &2   &1   &4  &3   \\
4            &1   &2   &3  &4   \\
\hline
\end{tabular}
\label{table:AlmostStable1}
\end{table}

\subsection{Two-Sided IA-Cell Matching}
In this section, we consider a different scenario in which \emph{low backhaul overhead is permitted before assignment}. By the GIA, each BS $k$ can compute $K-1$ \emph{potential} inner precoders $\{\mat{V}^{in}_{\ell}(k)\}_{\ell \neq k}$ for all the other cells based on $\{\mat{H}_{\ell}^k\}_{\ell \neq k}$, and then BS $k$ reports the potential inner precoders to the corresponding BSs via backhaul links, e.g., sending $\mat{V}_{k'}^{in}(k)$ to BS $k'$.

\subsubsection{{Preferences Generation}}
In this case, each cell not only knows the potential aligned interference subspace $\{\overline{\mat{F}}_{\ell}^k\}_{\ell \neq k}$ (corresponding to the potential IA-providers in the one-sided assignment) but also its \emph{potential} inner precoders $\{\mat{V}_k^{in}(k')\}_{k' \neq k}$ (corresponding to the potential IA-receivers). 
It is possible for each cell to compute double preferences for its IA-provider and IA-receiver. Let $\mathcal{P}_{k}^{p}$ and $\mathcal{P}_k^{r}$ be the  IA-provider preference list and  IA-receiver preference list, and both are \emph{incomplete preferences} with $K-1$ elements. 
More precisely, $\mathcal{P}_{k}^{p}$ is defined in \ref{eq:Preflist1}) and $\mathcal{P}_k^{r}$ can be generated by
\begin{align}
\mathcal{P}_k^{r} = &\mathrm{arglist}~\max_{\ell \neq k} \nonumber \\
&\sum_{i=1}^L\log_2\det\left( \mat{I}_{d_s} + \mat{V}_{i,k}^H(\ell)\mat{H}_{i,k}^{k,H} \mat{H}_{i,k}^k \mat{V}_{i,k}(\ell)\right),
 \label{eq:Preflist2}
\end{align}
where the performance metric has a "rate-like" form based on the available incomplete information.

\subsubsection{{Stable Marriage Model based IA-Cell Matching}}

In this two-sided IA-Cell matching, each cell hopes to find its most preferred IA-provider and IA-receiver, respectively. 
To balance the potential preference conflicts, the two-sided matching is required to determine a stable matching. In this case, the problem is well modeled by the well-known \emph{stable marriage matching with unacceptable partners} \cite{Gusfield1989} by considering each user group and BS as a man and a woman (or reversely), respectively. 
Based on \cite[Theorem 1.4.2]{Gusfield1989}, the following result holds. 
\begin{corollary}
Consider the strict IA-Cell assignment where user group $k$ and BS $k$ are unacceptable to each other. The stable matching may not exist (only one pair of user group and BS in a cell is not matched.) but is stable if it exists.  
\end{corollary}

To obtain the stable matching, following the same line of the one-sided matching, the strict two-sided IA-Cell assignment problem is first relaxed to a weak two-sided IA-Cell assignment problem. If the strict IA-Cell assignment has a stable matching, it can be efficiently determined 
by the basic Gale-Shapley algorithm \cite{Gale1962}. Otherwise, an "almost stable" matching can be obtained by a further breaking step.

We remark that an assignment by either the one-sided or two-sided stable matching scheme does not necessarily maximize the sum-cluster rate or the single-cell rate, since the goal is to find stable matchings and, additionally, only partial backhaul is used.

\subsection{{Centralized IA-Cell Assignment}}
Finally, we consider the case when \emph{there exists a central authority{\footnote{In the case of cellular networks this authority could be either a central controller (e.g., the Cloud-RAN) or a BS who serves as the cluster head and does the centralized optimization for the network. In particular, the cluster head could be a fixed or a rotating one.}} and high backhaul overhead is permitted}.
Without loss of generality, we assume BS $k$ serves as the cluster head and performs the assignment for all cells. Each BS $k', \forall k'\neq k$ sends the $K-1$ potential IA precoders $\{\mat{V}_{\ell}(k')\}_{\ell \neq k'}$ and the direct channel matrices $\mat{H}_{k'}^{k'}$ to BS $k$. Then, the optimal assignment for a certain problem, e.g., sum-cluster rate maximization or minimum single-cell rate maximization, can be determined by BS $k$ by brute-force search and based on the collected information. Afterwards, BS $k$ announces the assignment result to the cluster members. We stress that this rate  optimal assignment is not necessary to be stable.

\begin{remark}
From Lemma \ref{le:derangement}, there are few derangements for the cluster with a small number of cells, e.g., $2$ strict IA-Cell assignments for $K=3$ and $8$ strict IA-Cell assignments for $K=4$. In this case, the brute-force search is a reasonable approach.
 However, as $K$ increases, the number of derangements increases significantly, e.g., $264$ strict IA-Cell assignments for $K=6$, and the resulting backhaul overhead and the computational load become too large. 
\end{remark}

\section{Dynamic Feedback Bit Allocation under Limited Feedback} \label{sec:feedback}

Given an IA-Cell assignment, each BS $k$ obtains from its IA-provider its own IA precoder $\mat{V}_k^{in}$. Let $\overrightarrow{\mat{V}}_{i,k}\triangleq \mat{T}_{i}\mat{V}_{k}(\mat{V}_{k}^H\mat{T}_{i}^H\mat{T}_{i}\mat{V}_{k})^{-\frac{1}{2}}$ be the \emph{precoder pattern} in (\ref{eq:Vikoptimal}) where $\overrightarrow{\mat{V}}_{i,k}^H\overrightarrow{\mat{V}}_{i,k} = \mat{I}_{d_s}$. In order to implement a closed-loop transmission, $\overrightarrow{\mat{V}}_{i,k}$ needs to be fed back to user $(i, k)$. Since feedback links are usually capacity-limited, subspace quantization is employed to reduce overhead. A subspace matrix is mapped to an index in a predefined codebook. However, the use of a finite codebook inevitably causes a quantization distortion. As a result, perfect IA is no longer possible, and a residual interference term is to be managed. 
Therefore, the problem of DBA to minimize the sum-cluster RINR is of interest.

\subsection{Grassmannian subspace quantization}

Due to $\overrightarrow{\mat{V}}_{i,k}^H\overrightarrow{\mat{V}}_{i,k} = \mat{I}_{d_s}, \forall i, k$, subspace quantization can be applied to quantize the precoder patterns. 
Here, we give a subspace quantization example of a subspace matrix $\mat{V} \in \mathbb{C}^{M\times N}$ where $M > N$ by $B$ feedback bits. Assume that both BSs and users know the common codebook $\mathcal{C}$, i.e.,
\begin{align}
\mathcal{C} = \{\mat{C}_n \in \mathbb{C}^{M\times N}:~\mat{C}_n^H\mat{C}_n = \mat{I}_N, n=1, \ldots, 2^{B} \},
\end{align}
which can be generated and stored offline.
The quantized subspace is determined as the closest codeword in $\mathcal{C}$ by measuring the chordal distance
\begin{align}
\widehat{\mat{V}} &\triangleq \arg \min_{\mat{C}_n \in \mathcal{C}} d_c^2(\mat{V}, \mat{C}_{n})  \nonumber \\
&=  \arg \min_{\mat{C}_n \in \mathcal{C}} N- \trace(\mat{V}\mat{V}^H\mat{C}_n\mat{C}_n^H) \label{eq:Quan1}.
\end{align} 
The considered quantization is well-known as Grassmannian quantization on the Grassmann manifold $\mathcal{G}(M,N)$, defined as the set of the $N$-dimensional subspaces in the $M$-dimensional complex Euclidean space. Optimal Grassmann codebook designed based on \emph{Grassmannian subspace sphere-packing} is a challenging problem, which has attracted many research efforts \cite{Love2005, Ashikhmin2007, GrassmanSubspacePackingRobert, Schober2009, Medra2014} and references therein.

\begin{lemma}\label{le:QuanC} (Quantized Subspace Characterization)
The quantization $\widehat{\mat{V}} \in \mathbb{C}^{M\times N}$ of the subspace $\mat{V}\in \mathbb{C}^{M\times N}$ based on the subspace quantization can be characterized as
\begin{align}
\widehat{\mat{V}} = \mat{V}{\mat{R}} \mat{\Gamma}^{1/2} \mat{G}^H + \mat{V}^{\perp}{\mat{S}} (\mat{I}_N - \mat{\Gamma})^{1/2} \mat{G}^H \label{eq: QuanExpress}
\end{align}
where $\mat{V}^{\perp}\in\mathbb{C}^{M\times (M-N)}$ spans the left null space of $\mat{V}$,  
and $\mat{\Gamma} \triangleq \diag \{\alpha_1, \ldots, \alpha_N\}$ where $\alpha_j \in (0,1)$ and $\sum_{j=1}^N \alpha_j = N-d_c^2(\widehat{\mat{V}}, \mat{V})$, and $\mat{R} \in \mathbb{C}^{N\times N}$, $\mat{G} \in \mathbb{C}^{N\times N}$ and $\mat{S} \in \mathbb{C}^{(M-N) \times N}$ satisfy $\mat{R}^H\mat{R} = \mat{G}^H\mat{G}  = \mat{S}^H \mat{S} = \mat{I}_N$.
\end{lemma}
\begin{IEEEproof}
Please refer to the proof in Appendix \ref{subsec:ProofQuanC}.
\end{IEEEproof}

\begin{remark}
Since popular performance metrics, such as transmit power, minimum square error (MSE) and achievable rate, are functions of $\hat{\mat{V}}\hat{\mat{V}}^H$, the quantization characterization in (\ref{eq: QuanExpress}) can be further simplified to
\begin{align}
\widehat{\mat{V}} = \mat{V}{\mat{R}} \mat{\Gamma}^{1/2} + \mat{V}^{\perp}{\mat{S}} (\mat{I}_N - \mat{\Gamma})^{1/2}, \label{eq: QuanExpress2}
\end{align}
because $\widehat{\mat{V}}\widehat{\mat{V}}^H$ is independent of the unitary matrix ${\mat{G}}$ in (\ref{eq: QuanExpress}).
This quantized subspace characterization in (\ref{eq: QuanExpress2}) is more efficient than that in \cite[Lemma 1]{Ravindran2008} where $\mat{\Gamma}^{1/2}$ is an upper triangular matrix derived based on QR decomposition instead of a diagonal matrix as in our formulation.
\end{remark}

Based on a Grassmannian subspace sphere-packing codebook $\mathcal{C}$, the \emph{deterministic} subspace quantization distortion (\ref{eq:Quan1}) is defined by $d_c^2(\mat{V}, \widehat{\mat{V}}) \triangleq N - \trace(\mat{V}{\mat{V}}^H\widehat{\mat{V}}\widehat{\mat{V}}^H)$. Based on \cite[Theorem 4]{GrassmanSpherePackingbound1}, the maximum value of $d_c^2(\mat{V}, \widehat{\mat{V}})$ can be upper bounded by\footnote{For engineering purpose, this upper bound is obtained by omitting the $o(2^{-\frac{B}{2N(M-N)}})$ term in \cite[Theorem 4]{GrassmanSpherePackingbound1} for large codebooks due to $\lim_{B\rightarrow +\infty}2^{-\frac{B}{2N(M-N)}} \rightarrow 0$.}
\begin{align}
d_c^2(\mat{V}, \widehat{\mat{V}}) \leq \max_{\forall \mat{V} \in \mathcal{G}(M,N)} d_c^2(\mat{V}, \widehat{\mat{V}})  \leq  c(M,N) 2^{-\frac{B}{N(M-N)}}. \label{eq:Dis2}
\end{align}
In \eqref{eq:Dis2}, 
$c(M,N) \triangleq c^{-\frac{1}{N(M-N)}}$ is a constant coefficient, where $c$ is the coefficient of the metric ball volume of a subspace in the Grassmann manifold $\mathcal{G}(M,N)$ as specified in \cite[Eq. (8)]{Dai2008}.

\subsection{Dynamic IA Precoders Quantization and Feedback}\label{subsec: DFeed}
By the Grassmannian subspace quantization in (\ref{eq:Quan1}), each subspace matrix $\overrightarrow{\mat{V}}_{i,k}$ can be expressed by an index, which will be sent to user $(i, k)$ through the limited feedback link. Let $B_{i,k}$ denote the number of feedback bits for $\overrightarrow{\mat{V}}_{i,k}$ subject to a sum feedback bits constraint 
$\sum_{k=1}^K\sum_{i=1}^L B_{i,k} \leq B$.

Consider an IA-Cell assignment $Cell~k' \stackrel{IA}{\longrightarrow} Cell~k$. After subspace quantization and feedback of $\{\overrightarrow{\mat{V}}_{i,k'}\}_{i=1}^L$, the interference from cell $k'$ to cell $k$ with the quantized precoder pattern $\{\widehat{\mat{V}}_{i,k'}\}_{i=1}^L$, denoted by $\widehat{\mat{F}}_{k'}^k$, cannot be perfectly aligned into a $d_s$-dimensional subspace.  The imperfectly aligned interference spreads into a higher dimensional subspace, which cannot be completely removed by the ZF decoding. Thus, residual interference exists.

The total RINR from cell $k'$ to cell $k$ is defined as
\begin{align}
\mathcal{I}_{k'}^k \triangleq  \sum_{i=1}^L\mathcal{I}_{k'}^{i,k},
\end{align}
where $\mathcal{I}_{k'}^{i,k}$ denotes the RINR from cell $k'$ to user $(i,k)$, i.e.,
\begin{align}
\mathcal{I}_{k'}^{i,k} \triangleq \trace\left( \widehat{\mat{U}}_{i,k}^H \sum_{j=1}^L\frac{P_{j,k'}}{d_s\sigma_k^2}\left( \mat{H}_{j,k'}^{k} \widehat{\mat{V}}_{j,k'}\widehat{\mat{V}}_{j,k'}^H\mat{H}_{j,k'}^{k,H}  \right)\widehat{\mat{U}}_{i,k}\right),  \label{eq:ResiInF1}
\end{align}
where the decoder $\widehat{\mat{U}}_{i,k}$ is designed as
\begin{align}
\widehat{\mat{U}}_{i,k} \triangleq \left(\Big[\widehat{\mat{F}}_{j,k}^{IUI},~{\left\{\widehat{\mat{F}}_{\ell}^{k}\right\}_{\ell=1, \ell \neq k'}^K},~{\mat{H}_{i,k'}^{k} {\mat{V}}_{i,k'}^{in}} \Big] \right)^{\perp},   \label{eq: Uinhat}        
\end{align}
by which the interference from other cells $\ell \neq k'$ (except for the IA-provider cell $k'$) can be completely removed at BS $k$. 

Let $\mathcal{I}^k \triangleq \sum_{\ell=1}^K\mathcal{I}_{\ell}^k$ denote the total RINR from all cells to cell $k$, and thus we have $\mathcal{I}^k = \mathcal{I}_{k'}^k$ because of $\sum_{\ell \neq k'}\mathcal{I}_{\ell}^k = 0$ by the decoder \eqref{eq: Uinhat}.

\begin{proposition}\label{pr:InterferenceUpper}
Without loss of generality, under the IA-Cell assignment $Cell~k' \stackrel{IA}{\longrightarrow} Cell~k$,  
the total RINR to cell $k$ is upper bounded as 
\begin{align}
\mathcal{I}^{k} \leq \overline{\mathcal{I}}^{k} \triangleq L\overline{\mathcal{I}}_{k'}^{i,k},  \label{eq:upperInF0}
\end{align} 
 where  $\overline{\mathcal{I}}_{k'}^{i,k}$ denotes the upper bound of $ \mathcal{I}_{k'}^{i,k}$, i.e.,
\begin{align}
\overline{\mathcal{I}}_{k'}^{i,k}  \triangleq c(N_U, d_s)\sum_{j=1}^L \frac{P_{j,k'}}{\sigma_k^2 d_s}  \lambda_1(\mat{\Omega}_{j,k'}^{k})  2^{-\frac{B_{j,k'}}{ds(N_U-d_s)}}, \label{eq:upperInF}
\end{align}
with 
\begin{align}
\mat{\Omega}_{j,k'}^{k} \triangleq \left({\mat{V}}_{j,k'}^{in, \perp}\right)^H\mat{H}_{j,k'}^{k,H}\mat{\Pi}_{{\mat{H}}_{j,k'}^{k} \mat{V}_{j,k'}^{in}}^{\perp} \mat{H}_{j,k'}^{k} {\mat{V}}_{j,k'}^{in, \perp}.    \label{eq:Omegadef}
\end{align}
\end{proposition} 
\begin{IEEEproof}
Please refer to the proof in Appendix \ref{subsec:ProofIAupperbound}.
\end{IEEEproof}

In order to reduce the RINR, efficient usage of the limited feedback bits is desired.

\subsection{Dynamic Feedback Bit Allocation for Precoders} \label{sec:BitAllocation}

In this section, a DBA algorithm is studied to minimize the upper bound on the sum-cluster RINR.
\begin{equation}
 \begin{aligned}
\min_{\{\{B_{i,k}\}_{i=1}^L\}_{k=1}^K} ~~&\sum_{k=1}^K \overline{\mathcal{I}}^k  \\
\mathrm{s.t.}~~~~~&\sum_{k=1}^K\sum_{i=1}^L {B_{i,k}} \leq B; ~\forall B_{i,k} \in \mathbb{N}_0^+ \label{eq:SumMInbits}
 \end{aligned}
 \end{equation}
 where $\overline{\mathcal{I}}^k$ is given in \eqref{eq:upperInF0} and \eqref{eq:upperInF}. Observe that Problem (\ref{eq:SumMInbits}) is a jointly convex problem of $\{B_{i,k}\}$ when the non-negative integer constraint is relaxed and yields the following solutions.

Without loss of generality,  we hereafter assume that all the users transmit with the same uplink transmit power, i.e., $P_{i,k} = P, \forall i, k$ and all the BSs are with the same noise power, i.e., $\sigma_k^2 = \sigma^2, \forall k$, and define the transmit power to noise power ration (TSNR), i.e., $\SNR = \frac{P}{\sigma^2}$.

\begin{proposition}(Bit Allocation Solution) \label{pr:bitA}
Let us define 
\begin{align}
\mat{a} \triangleq \mathrm{arglist} \max_{\forall i; \forall k} \{\{\log_2 ( \lambda_1(\mat{\Omega}_{i,k}^{k+1}) )\}_{i=1}^L\}_{k=1}^K. \label{eq:lambdalist}
\end{align} 

Given an arbitrary $B$, the number of active users whose allocated feedback bit is positive can be determined by checking
\begin{align}
 \sum_{n=1}^{N_{a}} \mat{a}(n) - &N_a\mat{a}(N_{a})
\leq \frac{B}{d_s(N_U-d_s)}  \leq \nonumber \\
&~~~~~\sum_{n=1}^{N_{a}}\mat{a}(n) - N_a \mat{a}(N_a+1),  \label{eq:Bitcondition}
\end{align}
where $N_a \in \{1, \ldots, KL\}$ denotes the number of active users.
After determining $N_a$, the optimal solution for the $N_a$ active users in Problem (\ref{eq:SumMInbits})  is given in closed-form by
\begin{align}
B_{i,k}^{\star} = &\Big[d_s(N_U-d_s) \Big(\log_2(\lambda_1(\mat{\Omega}_{i,k}^{k+1})) -   \frac{1}{N_a}\sum_{n=1}^{N_a} \mat{a}(n) \nonumber \\
&~~~~~~~~~~~~~~~~~~~~~~+{\frac{B}{N_a d_s(N_U-d_s)}}\Big)\Big]_{int}.   \label{eq:Sumbits2}
\end{align}
And no feedback bits is allocated to those inactive users.
\end{proposition} 
\begin{IEEEproof}
The Lagrangian function with multiplier $\mu$ for Problem (\ref{eq:SumMInbits}) can be formulated as
\begin{align}
\mathcal{L}(\{\{B_{i,k}\}_{i=1}^L\}_{k=1}^K, \mu) = &\sum_{k=1}^K \sum_{i=1}^L \lambda_1(\mat{\Omega}_{i,k}^{k+1}) 2^{-\frac{B_{i,k}}{ds(N_U-d_s)}}  \nonumber \\
&+ \mu\Big(\sum_{k=1}^K \sum_{i=1}^L{B_{i,k}} - B\Big).
\end{align} 
With the definition $\zeta \triangleq \frac{d_s(N_U-d_s)}{\ln2}\mu$, the KKT conditions are
\begin{align}
&\frac{\partial{\mathcal{L}}}{\partial{B_k}} = -  \lambda_1(\mat{\Omega}_{i,k}^{k+1})2^{-\frac{B_{i,k}}{d_s(N_U-d_s)}} + \zeta = 0 \label{eq:Lagrangian1} \\
&\frac{\partial{\mathcal{L}}}{\partial{\zeta}} =  \sum_{k=1}^K\sum_{i=1}^L {B_{i,k}} - B = 0;~~\zeta > 0,  \label{eq:Lagrangian2}
\end{align} 
From (\ref{eq:Lagrangian1})-(\ref{eq:Lagrangian2}), we derive
\begin{align}
B_{i,k}(\zeta) = d_s(N_U-d_s) (\log_2( \lambda_1(\mat{\Omega}_{i,k}^{k+1}) )-\log_2(\zeta)),  \label{eq:Lagrangian3}
\end{align}
where $\zeta$ is determined such that $\sum_{k=1}^K\sum_{i=1}^LB_{i,k}(\zeta) = B$. Combining that $B_{i,k}$ is a nonnegative integer, we have
\begin{align}
B_{i,k}^{\star} = [ d_s(N_U-d_s) (\log_2( \lambda_1(\mat{\Omega}_{i,k}^{k+1}) ) - \log_2(\zeta) ) ]_{int}^+,  \label{eq:Sumbits1}
\end{align}
where $\zeta$ satisfies $\sum_{k=1}^K\sum_{i=1}^L B_{i,k}^{\star} = B$. 

To obtain the closed-form expression without variable $\zeta$, 
the water-filling principle implies that only the active users are allocated to the positive feedback bits. If there are $N_a$ active users where $N_a \in \{1, \ldots, KL\}$, with the definition in (\ref{eq:lambdalist}), the water-level satisfies
\begin{align}
\mat{a}(N_a+1) \leq \log_2(\zeta) \leq  \mat{a}(N_a). \label{eq:mu2} 
\end{align}
In the case of (\ref{eq:mu2}), plugging (\ref{eq:Sumbits1}) into (\ref{eq:Lagrangian2}) yields 
\begin{align}
\log_2(\zeta) &= \frac{1}{ N_a } \sum_{n=1}^{N_a} \mat{a}(n) -{\frac{B}{N_a d_s(N_U-d_s)}}. \label{eq:mu} 
\end{align}
Again plugging (\ref{eq:mu}) into (\ref{eq:Lagrangian3}) yelids (\ref{eq:Sumbits2}) under the condition (\ref{eq:Bitcondition}) that is obtained by combining (\ref{eq:mu}) and (\ref{eq:mu2}). There are $KL$ cases, i.e., $n \in \{1, \ldots, KL\}$. Given a $B$, we can determine how many and which users are active by checking (\ref{eq:Bitcondition}) and thus the closed-form bit allocation in (\ref{eq:Sumbits2}). 
\end{IEEEproof}

\subsection{Performance Analysis}
By treating residual interference as additive noise, we define the throughput under limited feedback of user $(i,k)$ as \cite{Rao2013}

\begin{align}
&\widehat{R}_{i,k} =  \log_2\det \Big(\mat{I}_{d_s} + \frac{\SNR}{d_s} \times \nonumber \\
&(\widehat{\mat{U}}_{i,k}^{H}\mat{H}_{i,k}^k\widehat{\mat{V}}_{i,k})(\widehat{\mat{U}}_{i,k}^H\mat{H}_{i,k}^k\widehat{\mat{V}}_{i,k})^H ( \mat{I}_{d_s} + \mat{C}_{i,k} )^{-1} \Big),  \label{eq:GeneralRate}
\end{align}
where $\mat{C}_{i,k} \triangleq \frac{\SNR}{d_s}\sum_{(j, \ell) \neq (i,k)} \widehat{\mat{U}}_{i,k}^H\mat{H}_{j,\ell}^k \widehat{\mat{V}}_{j,\ell}(\widehat{\mat{U}}_{i,k}^H\mat{H}_{j,\ell}^k\widehat{\mat{V}}_{j,\ell})^H$ denotes the overall residual interference matrix of user $(i,k)$. In the unlimited feedback case, 
(\ref{eq:GeneralRate}) is the same as (\ref{eq:RateIA1}). 

In order to further motivate the consideration of Problem \eqref{eq:SumMInbits}, we study the effect of sum feedback bit budget on the average sum cluster-rate under the IA-Cell assignment $Cell~k' \overset{IA}{\rightarrow} Cell~k$.

\begin{align}
&\widehat{R}_{sum}\triangleq \sum_{k=1}^K \sum_{i=1}^L \mathbb{E}(\widehat{R}_{i,k}) > \nonumber \\
&\sum_{k=1}^K \sum_{i=1}^L \mathbb{E}\Bigg(\log_2\Bigg(\frac{\trace\left( \widehat{\mat{U}}_{i,k}^{H}\mat{H}_{i,k}^k\widehat{\mat{V}}_{i,k}(\widehat{\mat{U}}_{i,k}^H\mat{H}_{i,k}^k\widehat{\mat{V}}_{i,k})^H\right)}{\frac{d_s}{\SNR}\trace\left( \mat{I}_{d_s} + \mat{C}_{i,k} \right)} \Bigg)\Bigg) \label{eq:Rsum1} \\
&\geq  \underbrace{\sum_{k=1}^K \sum_{i=1}^L \mathbb{E}(\log_2(\trace( \widehat{\mat{U}}_{i,k}^{H}\mat{H}_{i,k}^k\widehat{\mat{V}}_{i,k}(\widehat{\mat{U}}_{i,k}^H\mat{H}_{i,k}^k\widehat{\mat{V}}_{i,k})^H)) )}_{\triangleq \overline{R}_{sum}}  \nonumber \\
&~~~~~~~~~~-\sum_{k=1}^K \sum_{i=1}^L\mathbb{E}\Big(\log_2\Big(\frac{d_s}{\SNR}( d_s +  \overline{\mathcal{I}}_{k'}^{i,k} )\Big)\Big)  
 \label{eq:Rsum2} \\
&\approx \overline{R}_{sum} -  \sum_{k=1}^K \sum_{i=1}^L\mathbb{E}\Big(\log_2\Big(\frac{d_s}{\SNR}\overline{\mathcal{I}}_{k'}^{i,k} \Big)\Big)     \label{eq:Rsum4}  \\
&= \overline{R}'_{sum} - \sum_{k=1}^K\sum_{i=1}^L\mathbb{E}\Big(\log_2\Big(\sum_{j=1}^L\lambda_1(\mat{\Omega}_{j,k'}^{k})  2^{-\frac{B_{j,k'}}{ds(N_U-d_s)}} \Big)\Big)  
 \label{eq:Rsum5} \\
&\geq \overline{R}'_{sum} - \sum_{k=1}^K\sum_{i=1}^L\mathbb{E}\Big(\log_2\Big(2^{-\frac{B_{i,k'}}{ds(N_U-d_s)}} \sum_{j=1}^L\lambda_1(\mat{\Omega}_{j,k'}^{k}) \Big)\Big)  
 \label{eq:Rsum6} \\
&=\overline{R}''_{sum} -\mathbb{E}\Big(\log_2\Big( \prod_{k=1}^K\prod_{i=1}^L 2^{-\frac{{B}_{i,k}}{ds(N_U-d_s)}} \Big)\Big)  
 \label{eq:Rsum7} \\
&=\overline{R}_{sum} + \frac{1}{d_s(N_U-d_s)}B   \label{eq:Rsum8}
\end{align}
where the inequality in (\ref{eq:Rsum1}) is based on \cite[Theorem 1]{Sun2012} and $\log(1+x) > \log(x)$, and the inequality in (\ref{eq:Rsum2}) is based on $\trace(\mat{C}_{i,k}) = \mathcal{I}_{k'}^{i,k} \leq  \overline{\mathcal{I}}_{k'}^{i,k}$, where $\overline{\mathcal{I}}_{k'}^{i,k}$ is the RINR upper bound of user $(i, k)$ shown in \eqref{eq:upperInF}.
 The approximation \eqref{eq:Rsum4} is under the assumption $\overline{\mathcal{I}}_{i,k} \gg d_s$ (we will discuss this assumption in the following). 
Based on (\ref{eq:upperInF}) and the definition of $\overline{R}'_{sum} \triangleq \overline{R}_{sum} - KL\log_{2}(c(N_U, d_s))$, we equivalently have (\ref{eq:Rsum5}). Under the assumption $B_{1, k'} =\ldots =B_{L, k'}$ (i.e., equal feedback bits among the users within each cell), (\ref{eq:Rsum6}) surely serves as an lower bound of the DBA. Equation \eqref{eq:Rsum6} is obtained based on the definition of $\overline{R}''_{sum} \triangleq \overline{R}'_{sum} -  \sum_{k=1}^K\sum_{i=1}^L\mathbb{E}\Big(\log_2\Big(\sum_{j=1}^L\lambda_1(\mat{\Omega}_{j,k'}^{k}) \Big)\Big)$. Finally, \eqref{eq:Rsum8} holds recalling that $\sum_{k=1}^K\sum_{i=1}^L B_{i,k} = B$.

\begin{remark}
From \eqref{eq:Rsum4}, we observe that the lower bound of the average sum cluster-rate is approximately decreasing  with RINR, which implies that it is reasonable to design the feedback bit allocation policy to suppress the residual interference, as our formulated problem \eqref{eq:SumMInbits}.

From \eqref{eq:Rsum8}, recalling the expression of $\overline{R}''_{sum}$, we observe that in $\overline{R}''_{sum}$ only the term $\overline{R}_{sum}$ is related to the feedback bits, because the quantized precoder $\widehat{\mat{V}}_{i,k}$ is a combination of $\mat{V}_{i,k}$ and $\mat{V}_{i,k}^{\perp}$ with different weights (related to $B$). However, the components $\mat{V}_{i,k}$ and $\mat{V}_{i,k}^{\perp}$ of $\widehat{\mat{V}}_{i,k}$ in (\ref{eq: QuanExpress2}) are \emph{isotropic} and have the same effect in probability on $\mat{H}_{i,k}^k$, since ${\mat{V}}_{i,k}$ and also ${\mat{U}}_{i,k}$ are designed independently of $\mat{H}_{i,k}^k$. Therefore, $B$ has a slight influence on $\overline{R}_{sum}$ and thus $\overline{R}''_{sum}$.
In this case, the proposed lower bound of the average sum-cluster rate is linearly scaled by the third term with the rate of $\frac{1}{d_s(N_U-d_s)}$.
\end{remark}

\emph{Discussion on the Assumption of $\overline{\mathcal{I}}_{k'}^{i,k} \gg d_s $:} This assumption is equivalent to $\overline{\mathcal{I}}_{k'}^{i,k} \geq \rho d_s$ where $\rho$ is a scalar much larger than one\footnote{
By $\overline{\mathcal{I}}_{i,k} = \rho d_s$, we have $\log_2(d_s + \overline{\mathcal{I}}_{i,k}) = \log_2((1+\rho)d_s)$. In oder to measure the accuracy of the approximation of $\log_2((1+\rho)d_s)\approx \log_2(\rho d_s)$. 
Define
$\eta \triangleq \frac{\log_2(\rho d_s)}{\log_2((1+\rho)d_s)}  = \frac{\log_2(\rho) + \log_2( d_s)}{\log_2(1+\rho) + \log_2(d_s)}  {\geq}  \frac{\log_2(\rho)}{\log_2(1+\rho)}$. 
Therefore, it is sufficient to determine the value of $\rho$ such that 
$\eta > \frac{\log_2(\rho)}{\log_2(1+\rho)} \approx 1$, e.g., $\eta >  0.9900$ and  $\eta >  0.9978$ for $\rho =29$ and $\rho = 100$, respectively.}. 
By (\ref{eq:upperInF}), it is equivalent to 
\begin{align}
& c(N_U, d_s)\frac{\SNR}{d_s} \sum_{j=1}^L \lambda_1(\mat{\Omega}_{j,k'}^{k})  2^{-\frac{B_{j,k'}}{ds(N_U-d_s)}}  \geq \rho d_s  \label{eq: condB1} \\
&\Leftrightarrow~ c(N_U, d_s)\frac{\SNR}{d_s} L\zeta  \geq \rho d_s \label{eq: condB2} \\
&\Leftrightarrow~ \log_2(\zeta)  \geq \log_2\Big(\frac{\rho d_s^2}{Lc(N_U, d_s){\SNR}}\Big)  \label{eq: condB3} \\
&\Leftrightarrow~B \leq d_s(N_U-d_s)\times \nonumber \\ 
&~~~~~~~\Big(\sum_{n=1}^{N_a}\mat{a}(n) - N_a\log_2\Big(\frac{\rho d_s^2}{Lc(N_U, d_s){\SNR}}\Big) \Big), \label{eq: condB4}
\end{align}
where  (\ref{eq: condB2}) is based on (\ref{eq:Lagrangian1}), since the feedback bits are allocated based on Proposition \ref{pr:bitA}.
Plugging (\ref{eq:mu}) into (\ref{eq: condB3}) yields (\ref{eq: condB4}).  
Therefore, combining (\ref{eq:Bitcondition}) and (\ref{eq: condB4}), we have 
\begin{align}
\SNR \geq \frac{\rho d_s^2}{Lc(N_U, d_s)2^{\mat{a}(N_a)}},
\end{align}
which implies that the assumption $\overline{\mathcal{I}}_{k'}^{i,k} \gg d_s $ has different SNR requirements for different scenarios.

\section{Implementation and Analysis} \label{sec:analysis}

In this section, we analyze the following aspects of the proposed algorithm: {1})  implementation,  {2})  required overhead  and 3) complexity. 
\subsection{Implementation}
The outline of the implementation of the proposed algorithm is shown as follows, where each step could be a time slot. 
\begin{itemize}
\item \textbf{Step 1 ({CSIR estimation})}: Each BS $k$ estimates its local CSIR $\{\mat{H}_{\ell}^k\}_{\ell=1}^K$ based on orthogonal uplink pilot signals;
\item \textbf{Step 2 ({IA percoder computation})}:  Each BS $k$ employs the GIA method to compute $K-1$ \emph{potential} IA precoders $\{\mat{V}_{\ell}^{in}(k)\}_{\ell=1, \ell \neq k}^K$ for $K-1$ cells based on $\{\mat{H}_{\ell}^k\}_{\ell=1, \ell \neq k}^K$;
\item \textbf{Step 3 ({IA-Cell assignment})}: A suitable IA-Cell assignment is chosen from the following three schemes for  the considered system configuration.
\begin{itemize}
\item \emph{With no Backhaul Overhead Before Assignment (Distributed):} Based on only $\{\mat{V}_{\ell}^{in}(k)\}_{\ell=1, \ell \neq k}^K$ at each BS $k$, one-sided matching is implemented;
\item \emph{With low Backhaul Overhead Before Assignment (Distributed):} Each BS $k$ reports its computed $\{\mat{V}_{\ell}^{in}(k)\}_{\ell=1, \ell \neq k}^K$ to the $K-1$ corresponding BSs. Based on the collected IA precoders and its local CSIR, two-sided matching is implemented;
\item \emph{With high Backhaul Overhead Before Assignment (Centralized):} Assume that BS $k$ is the cluster head. Each BS $k' \neq k$ reports its computed $\{\mat{V}_{\ell}^{in}(k')\}_{\ell=1, \ell \neq k'}^K$ and its direct channels $\mat{H}_{k'}^{k}$ to the cluster head BS $k$ via backhaul links. Based on the collected informations, BS $k$ finds the optimal assignment by brute force search and communicates the assignment to each cell;
\end{itemize}

Once a good IA-Cell assignment is found by the chosen IA-Cell assignment scheme, its corresponding assigned perfect IA precoders and decoders can be determined. 
\item \textbf{Step 4 ({DBA})}:  
After determining the perfect IA transceivers for a given IA-Cell assignment, each BS $k$ needs to feed back its IA precoder patterns $\mat{V}_{i,k}$ to its users.
In order to enable efficient feedback of $\{\{\overrightarrow{\mat{V}}_{i,k}\}_{i=1}^L\}_{k=1}^K$, the DBA  is performed and yields the solution $\{\{B_{i,k}\}_{i=1}^L\}_{k=1}^K$ for the quantization of $KL$ precoder patterns; 
\item \textbf{Step 5 ({Quantization under limited feedback})}: Each BS $k$ quantizes the precoder patterns $\{\overrightarrow{\mat{V}}_{i,k}\}_{i=1}^L$ to $\{\widehat{\mat{V}}_{i,k}\}_{i=1}^L$ by Grassmannian subspace codebooks with size $\{2^{B_{i,k}}\}_{i=1}^L$ and broadcasts the indexes to its users; 
\item \textbf{Step 6 ({Uplink transmission})}: Based on the received index, each user $(i, k)$ selects the corresponding codeword from the codebook, i.e., $\widehat{\mat{V}}_{i,k}$, as its IA precoder pattern. Then, the quantized uplink precoder designed by \eqref{eq:Vik2} will be used for uplink transmission during the whole coherence time period.
\end{itemize}

\subsection{Backhaul overhead}

The required backhaul overhead (excluding the feedback overhead) of the different IA-Cell assignment schemes  are reported in  
Table \ref{table:overhead}, where "One-sided"/"Two-sided"/"Centralized"/"Fixed" denotes that one-sided/two-sided/centralized/fixed matching is used.

\begin{table*}
\centering
\begin{threeparttable}
\caption{Total backhaul overhead of  $K$ cells}\label{table:overhead}
\begin{tabular}{l|c|c|c}
\hline
Algorithms & Before assignment   &Assignment   &After assignment \\
\hline
One-sided         &0   &$4(K+(N_C-1))$ bit  &$KLN_Ud_s$ cc $+ (K-1)B$ bit   \\
Two-sided              &$K(K-1)LN_Ud_s$ cc  &$4[K, K^2-K+1]$ bit  & $(K-1)B$ bit   \\
Centralized              & $\substack{(K-1)^2LN_Ud_s +\\ (K-1)LN_UN_B}$ cc  &0  &$(K-1)LN_Ud_s$ cc $+ (K-1)B$ bit   \\
Fixed             &0   &--  &$KLN_Ud_s$ cc \\
\hline
\end{tabular}
\begin{tablenotes}
      \small
       \item 1) "cc" denotes the unit of a complex coefficient. 2) Each ask is responsed during the assignment.
    \end{tablenotes}
  \end{threeparttable}
\end{table*}

During the IA-Cell assignment by the one/two-sided matching, each BS $k$ has four possible actions to other BSs, namely "ask", "definitely accept", "temporarily accept" and "definitely reject", which can be encoded into two bits.
In particular, the one-sided matching by the FCA takes $K+(N_C-1)$ steps where $N_C$ denotes the number of cycle chains, and each step has one "ask" action.
The two-sided matching by the Basic Gale-Shapley algorithm \cite{Gale1962} takes $[K, K(K-1)+1]$ proposals. 
After assignment by the one-sided matching, each cell needs to send an explicit inner precoder to its corresponding IA-provider, while it is not necessary for the two-sided matching because it has been already exchanged before assignment. After the quantization of the precoder patterns, each BS needs to exchange the corresponding indexes with other BSs, based on which the new ZF decoder can be designed. 
The resulting total backhaul overhead is reported in Table \ref{table:overhead}.

\subsection{Complexity}

As shown in Section \ref{sec:ICA}, the complexity of computing $K$ IA precoders by the GIA is $K\mathcal{O}((L-1)^2LN_B^2 N_U)$. 

For the one-sided matching, the complexity mainly depends on the preference generation (\ref{eq:Preflist1}). 
 The generation of $K$ \emph{ranked} preference lists takes $K(K-1)L(\mathcal{O}(N_BN_Ud_s) +  2L(\mathcal{O}(N_U N_B^2) + \mathcal{O}(N_U^3)) + 2\mathcal{O}(N_B d_s^2) + 2\mathcal{O}(d_s^3) + K\mathcal{O}(K))$ arithmetic operations. The FCA with $K+(N_C-1)$ steps has complexity $\mathcal{O}(K)$ where $N_C$ denotes the number of cycle chains.
For the two-sided matching, 
besides generating (\ref{eq:Preflist1}),
 $K$ \emph{ranked} preference lists generation as in (\ref{eq:Preflist2}) requires $K(K-1)L\left(\mathcal{O}(2N_U^2d_s) +  \mathcal{O}(N_UN_B d_s) + 2(\mathcal{O}(d_s^3)\right) + K\mathcal{O}(K)$ arithmetic operations. The complexity of the Basic Gale-Shapley algorithm with at most $K^2 -K +1$ steps is upper bounded by $\mathcal{O}(K^2)$.   
The centralized assignment needs to compute $K!\sum_{k=0}^K\frac{(-1)^k}{k!}-1$ possible rate performance with complexity $(K!\sum_{k=0}^K\frac{(-1)^k}{k!}-1)K(L(\mathcal{O}(2N_U^2 d_s) + \mathcal{O}(N_B^2d_s ) + (L+1)\mathcal{O}(N_BN_Ud_s) + (L+2)\mathcal{O}(d_s^3) + (L+2)\mathcal{O}(N_B d_s^2)))$.

\begin{figure}[t]
         \centering
           \includegraphics[width=.4\textwidth]{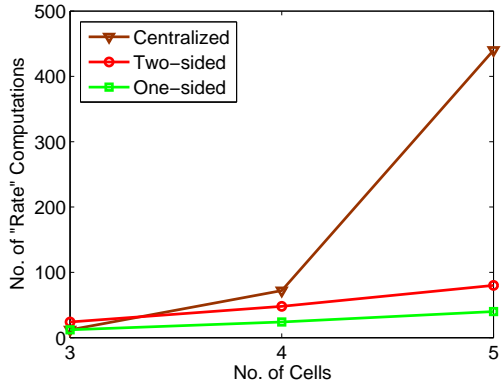}
                    \caption{Complexity comparison of the stable matching and centralized assignment}\label{fig:Complexity}
   \end{figure}

Roughly speaking, the one-sided matching, the two-sided matching and the centralized assignment mainly take $K(K-1)L$, $2K(K-1)L$ and $K!\sum_{k=0}^K\frac{(-1)^k}{k!}-1$ "rate-like" computations{\footnote{The computation expression is not the actual rate expression, but has always the form $\log_2\det(\mat{I} + \mat{X}\mat{\Pi}_{\mat{Y}}^{\perp}\mat{X}^H)$.}} , respectively. 
Fig. \ref{fig:Complexity} shows the approximate complexity of these three algorithms over the number of cells. It implies the centralized assignment is a reasonable approach with a comparable complexity as the distributed algorithms if $K\leq 4$. Instead, when $K\geq 5$ distributed algorithms are preferable as far as complexity is concerned. 

\section{Numerical Results}\label{sec:simulations}

In this section, the performance of the GIA with optimized IA-Cell assignment with both unlimited and limited feedback is evaluated. 
\subsection{System Model and Performance Metrics}
We consider a $(K, L, N_B, N_U, d_s) = (4, 2, 14, 8, 2)$ interfering MIMO-MAC. We set $\sigma_k^2 =1, \forall k$ and $P_{i,k}=P, \forall i, k$, respectively.
Let $\SNR = 10\log_{10}(P)$ denote the TSNR in dB.
 The path loss of direct links is set to be $1$, whereas the path loss of the cross links
is uniformly distributed in $[0,1]$, respectively{\footnote{This is to guarantee that interference channels are not stronger than direct channels, since a user is usually assigned to the BS who provides it the strongest link. The user selection and user-BS association can be done based on the uplink CSI available at BSs, which is out of the scope of this work.
 }}.
 
To properly measure the performance of the proposed approaches, we consider two following metrics
\begin{align}
R_{sum} \triangleq  \mathbb{E}\Big(\sum_{k=1}^K\sum_{i=1}^L\widehat{R}_{i,k} \Big), R_{min} \triangleq \mathbb{E}\Big( \min_{k=1,\ldots, K} \sum_{i=1}^L \widehat{R}_{i,k} \Big), \nonumber
\end{align}
where $\log_e(\cdot)$ is used in the rate expression of $\widehat{R}_{i,k}$ in (\ref{eq:GeneralRate}). $R_{sum}$ and $R_{min}$ are the average sum-cluster rate and the average minimum single-cell rate over different channel realizations. These performance functions measure the overall cluster throughput and the fairness of the cluster, respectively.

\subsection{Performance Comparison with Unlimited Feedback}
Under unlimited feedback, the effect of IA-Cell assignment on $R_{sum}$ and  $R_{min}$ is evaluated by the following metrics.
\begin{itemize}
    \item $\mathrm{Upper_{sum}}$ and $\mathrm{Lower_{sum}}$ ($\mathrm{Upper_{min}}$ and $\mathrm{Lower_{min}}$) denote the performance achieved by \emph{the best} and \emph{the worst} IA-Cell assignment for \emph{sum cluster-rate maximization} (\emph{minimum cluster-rate maximization}), respectively, which are determined by the centralized assignment;
    \item $\mathrm{Two}$/$\mathrm{One}$/$\mathrm{Fixed}$: Each channel realization is under the IA-Cell assignment by the two-sided/one-sided/fixed matching (\ref{eq:cellIASeassign});  
     \item $\mathrm{RB}$: Each precoder $\overrightarrow{\mat{V}}_{i,k}$ is a random subspace and each decoder is the "matched filter" $\mat{U}_{i,k} = \mat{H}_{i,k}^k\mat{V}_{i,k}(\mat{V}_{i,k}^H\mat{H}_{i,k}^{k,H}\mat{H}_{i,k}^k\mat{V}_{i,k})^{-\frac{1}{2}}$;
     \item $\mathrm{FDMA}$: Each user ocuppies an un-overlapped spectrum. 
\end{itemize}

\begin{figure}[t]
         \centering
           \includegraphics[width=.45\textwidth]{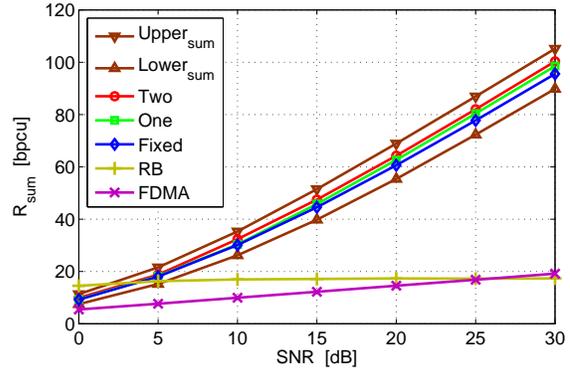}
                    \caption{Sum-cluster rate comparison under unlimited feedback w.r.t. SNR.}\label{fig:SumratePerfectCSI}
   \end{figure}

\begin{figure}[t]
                 \centering
           \includegraphics[width=.45\textwidth]{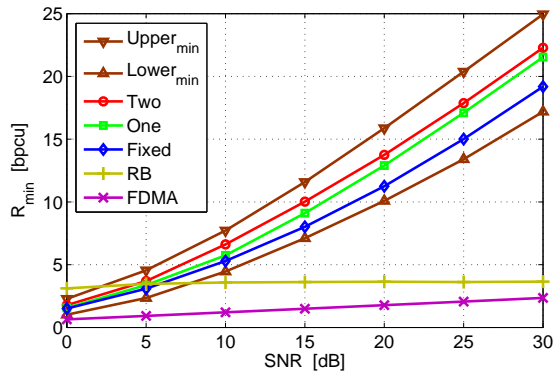}
            \caption{Minimum single-cell rate comparison under unlimited feedback w.r.t. SNR.}\label{fig:MinratePerfectCSI}
   \end{figure}
Both Fig. \ref{fig:SumratePerfectCSI} and Fig. \ref{fig:MinratePerfectCSI} show that a large performance gap exists between the best IA-Cell assignment and the worst IA-Cell assignment. It implies the IA-Cell assignment has a significant influence on both the overall throughput and the fairness. This performance gap regarding $R_{sum}$ is as large as $5$ dB and that of $R_{min}$ is even larger than $10$ dB for high SNR. Compared with the fixed matching, the two-sided and one-sided matching have a similar performance improvement, i.e., more than $1$ dB for $R_{sum}$ and more than $5$ dB for $R_{min}$. In Fig.  \ref{fig:SumratePerfectCSI}, it is observed that the sum cluster-rate curves by different strict IA-Cell assignments have different rate performance but \emph{in parallel} (with the same slope), which coincides with our theoretical analysis that different strict IA-Cell assignments yield different sum cluster-rate performance but the same DoFs (Corollary \ref{co: assignmentDoFs}).
The advantage of the GIA is obvious and significant compared with the random beamforming and FDMA, especially for high SNR.  

\subsection{Performance Comparison under Limited Feedback}
Under limited feedback, the proposed DBA is evaluated by comparing with the classical EBA (plotted in dashed lines in the following figures). The theoretical analysis of subspace quantization is based on the Grassmannian sphere-packing codebook. However, since it is extremely difficult to construct  large codebooks based on good Grassmannian sphere-packings, random subspace codebooks are adopted in the simulation\footnote{Note that the performance by random subspace codebooks constitutes a lower bounder to the performance by sphere-packing codebooks. In fact, for large codebooks, random subspace codebooks usually attain a similar numerical performance to sphere-packing codebooks, e.g., \cite{Rezaee2012}.}.

\subsubsection{Performance comparison w.r.t. sum feedback bit budget}

\begin{figure}[t]
         \centering
           \includegraphics[width=.45\textwidth]{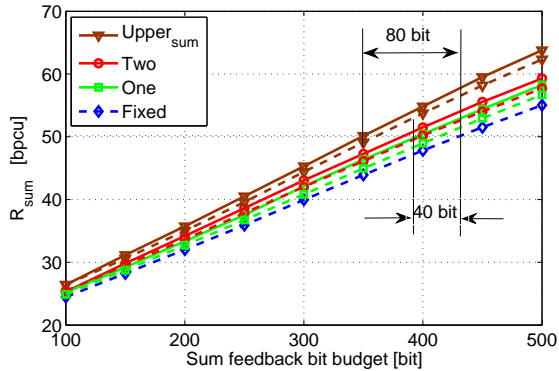}
                    \caption{Sum-cluster rate comparison under limited feedback w.r.t. sum feedback bit budget.}\label{fig:SumrateLFbit}
  \end{figure}

\begin{figure}[t]
                 \centering
           \includegraphics[width=0.45\textwidth]{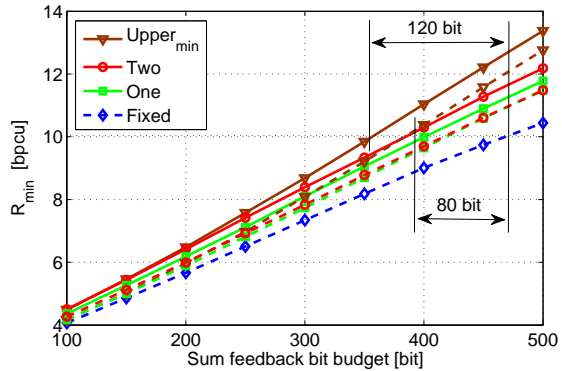}
            \caption{Minimum single-cell rate comparison under limited feedback w.r.t. sum feedback bit budget.}\label{fig:MinrateLFbit}
   \end{figure}

\begin{figure}[t]
                 \centering
           \includegraphics[width=.45\textwidth]{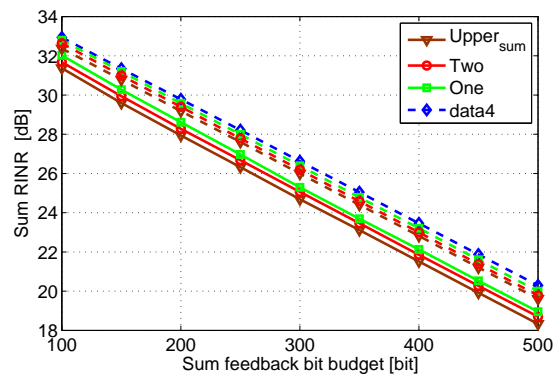}
            \caption{Sum-cluster RINR comparison under limited feedback w.r.t. sum feedback bit budget.}\label{fig:RINRbit}
   \end{figure}

The performance w.r.t. the sum feedback bit budget is evaluated when $\SNR=25$ dB. 
Fig. \ref{fig:SumrateLFbit} shows that the sum-cluster rate is increasing with the sum feedback bit budget \emph{at an approximate linear rate} of $0.09$, which approximately coincides with $\frac{1}{\log_2(e)}\frac{1}{d_s(N_U -d_s)}=0.0577$ in (\ref{eq:Rsum8}). The proposed DBA outperforms the EBA in both the sum cluster-rate in Fig. \ref{fig:SumrateLFbit} and the minimum single-cell rate in Fig. \ref{fig:MinrateLFbit}. Compared with the fixed matching with the EBA, the proposed centralized assignment and the distributed stable matching with the DBA can save around $80$ bit and $40$ bit, respectively, to achieve $R_{sum} = 50$ bpcu in Fig. \ref{fig:SumrateLFbit}, and around $120$ bit and $80$ bit, respectively, to achieve $R_{min} = 10$ bpcu in Fig. \ref{fig:MinrateLFbit}. 
The sum-cluster RINR in $10\log_{10}(\sum_{\ell=1}^K\mathcal{I}^k)$ dB is linearly decreasing with sum feedback bit budget as shown in Fig. \ref{fig:RINRbit}. The DBA achieves a lower RINR compared with the EBA, which implies that the effectiveness of minimizing the upper bound of sum-cluster RINR in (\ref{eq:upperInF}). The sum-cluster RINR is greatly larger than $d_s$ in Fig. \ref{fig:RINRbit}, making the approximation in (\ref{eq:Rsum5}) feasible. 

\subsubsection{Performance comparison w.r.t. SNR}

The proposed algorithms are evaluated by measuring the sum-cluster rate and the single-cell rate performance w.r.t. SNR for the fixed sum feedback budget $B=300$ bit and $B=500$ bit, respectively. 

\begin{figure}[t]
         \centering
           \includegraphics[width=.45\textwidth]{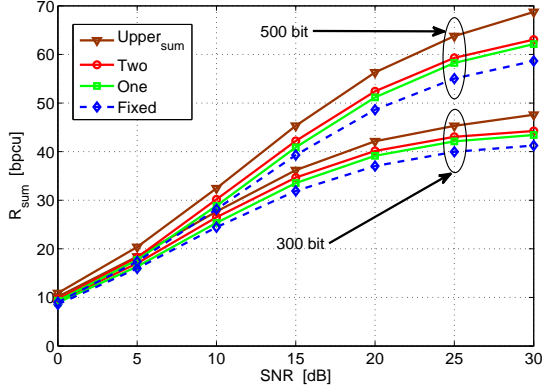}
                    \caption{Sum cluster-rate comparison under limited feedback w.r.t. SNR.}\label{fig:SumrateLFSNR}
  \end{figure}

\begin{figure}[t]
                 \centering
           \includegraphics[width=.45\textwidth]{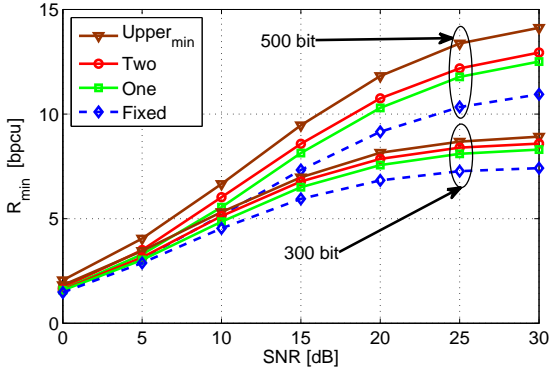}
            \caption{Minimum single-cell rate comparison under limited feedback w.r.t. SNR.}\label{fig:MinrateLFSNR}
   \end{figure}

From Fig. \ref{fig:SumrateLFSNR} and Fig. \ref{fig:MinrateLFSNR}, it is observed that the performance with $B=500$ bits is significantly better than that with $B=300$ bits and the performance gap enlarges with the SNR. For $\SNR=30$ dB, the gap of sum-cluster rate and that of the single-cell rate are as large as around $20$ bpcu and $8$ bpcu, respectively. From the perspective of energy consumption, the feedback of $B=500$ bits results in a higher complexity and more feedback energy consumption than the feedback of $B=300$ bits, while it is still attractive when battery power saving is the goal. This feature is very useful since the user terminals' battery power can be saved at the expense of a larger energy consumption at the BSs, where the virtually unlimited energy supply of the electric grid is available.
For example, $15$ dB uplink power can be saved by the stable matching to achieve $R_{sum}=40$ bpcu with $B=500$ bits compared with $B=300$ bits. Compared to the fixed matching with EBA, the proposed centralized assignment and stable matching with DBA can reduce by $10$ dB and $5$ dB uplink power, respectively, at an achieved rate of $R_{sum}=60$ bpcu. And this performance improvement enlarges with SNR.    

   \section{Conculsions}
 In this work, we provide a framework for the GIA with optimized IA-Cell assignment in the interfering MIMO MAC network under limited feedback. This algorithm 
 yields the closed-form IA transceiver by distributed implementation at the BS side if its feasible conditions are satisfied. Furthermore, the effect of IA-Cell assignment and DBA on either the sum-cluster rate or minimum single-cell rate are discussed and illustrated in the simulations.

\appendices

\section{Proof of Lemma \ref{le:tight}} \label{subsec:Prooftight}
\begin{IEEEproof}
Without loss of generality, let us consider an GIA example of $Cell~k \overset{IA}{\rightarrow} Cell~k+1$ to align $\{\mat{H}_{i,k}^{k+1}\mat{V}_{i,k}\} \in \mathbb{C}^{N_B \times d_s},~\forall i =1, \ldots, L$ by designing the precoders $\{\mat{V}_{i,k}\} \in  \mathbb{C}^{N_U \times d_s}$ subject to transmit power constraints $\trace(\mat{V}_{i,k}\mat{V}_{i,k}^H) \leq P_{i,k}$, the GIA demands 
\[
\begin{cases}
\mathrm{Span}(\mat{H}_{1,k}^{k+1}\mat{V}_{1,k}) = \ldots = \mathrm{Span}(\mat{H}_{L,k}^{k+1}\mat{V}_{L,k}) \\
\trace(\mat{V}_{i,k}\mat{V}_{i,k}^H) \leq P_{i,k},~\forall i = 1, \ldots, L
\end{cases}
\]
\[
\overset{(a)}{\Leftrightarrow}
\begin{cases}
\mathrm{Span}(\mat{H}_{1,k}^{k+1}\mat{Q}_{1,k}) =\ldots = \mathrm{Span}(\mat{H}_{L,k}^{k+1}\mat{Q}_{L,k}) ~~~~~\\
\trace(\mat{R}_{i,k}\mat{R}_{i,k}^H) \leq P_{i,k},~\forall i = 1, \ldots, L
\end{cases}
\]
\[
\overset{(b)}{\Leftrightarrow}
\begin{cases}
\mat{H}_{1,k}^{k+1}\mat{Q}_{1,k}\tilde{\mat{X}}_{1} = \ldots = \mat{H}_{L,k}^{k+1}\mat{Q}_{L,k}\tilde{\mat{X}}_{L} ~~~~~~~~~~~~~~ \\
\trace(\mat{R}_{i,k}\mat{R}_{i,k}^H) \leq P_{i,k},~\forall i = 1, \ldots, L
\end{cases}
\]
\[
\overset{(c)}{\Leftrightarrow}
\begin{cases}
\mat{H}_{1,k}^{k+1}{\mat{V}}_{1,k}^{in} = \ldots = \mat{H}_{L,k}^{k+1}{\mat{V}}_{L,k}^{in}  \\
 \trace({\mat{V}}_{i,k}^{in}{\mat{V}}_{i,k}^{out}{\mat{V}}_{i,k}^{out,H}{\mat{V}}_{i,k}^{in, H}) \leq P_{i,k},~\forall i = 1, \ldots, L
\end{cases}
\]
where (a) is based on the QR decomposition of  $\mat{V}_{i,k} \triangleq \mat{Q}_{i,k}\mat{R}_{i,k}$ and $\mathrm{Span}(\mat{H}_{i,k}^{k+1}\mat{Q}_{i,k}\mat{R}_{i,k}) = \mathrm{Span}(\mat{H}_{i,k}^{k+1}\mat{Q}_{i,k})$, where $\mat{Q}_{i,k}\in\mathbb{C}^{N_{U} \times d_s}$ and $\mat{R}_{i,k}\in\mathbb{C}^{d_s \times d_s}$ denote the ''subspace'' and the ''power'' of $\mat{V}_{i,k}$, respectively. Based on the equivalence (a), $\{\mat{V}_{i,k}\}$ can be determined  via  independently designing $\{\mat{Q}_{i,k}\}$ based on the IA constraint  and $\{\mat{R}_{i,k}\}$ subject to the power constraints. The equivalence (b) is because the restriction of the IA condition has an influence only on the "power" of $\mat{H}_{i,k}^{k+1}\mat{Q}_{i,k}\tilde{\mat{X}}_{i}$ by introducing a full rank matrix $\tilde{\mat{X}}_i \in \mathbb{C}^{d_s \times d_s}$ but not on the its "subspace", where $\{\tilde{\mat{X}}_i\}$ are selected to fulfill $\mat{H}_{1,k}^{k+1}\mat{Q}_{1,k}\tilde{\mat{X}}_{1} = \ldots = \mat{H}_{L,k}^{k+1}\mat{Q}_{L,k}\tilde{\mat{X}}_{L}$. Therefore, $\mat{H}_{1,k}^{k+1}\mat{Q}_{1,k}\tilde{\mat{X}}_1 = \ldots = \mat{H}_{L,k}^{k+1}\mat{Q}_{L,k}\tilde{\mat{X}}_L$ is a \emph{necessary but not sufficient} condition of $\mathrm{Span}(\mat{H}_{1,k}^{k+1}\mat{Q}_{1,k}) = \ldots =\mathrm{Span}(\mat{H}_{L,k}^{k+1}\mat{Q}_{L,k})$ in terms of the "power" because of the restriction on $\tilde{\mat{X}}_i$ (in fact, this "power" restriction can be completely eliminated when the transmit power constraints are jointly considered), and they are \emph{equivalent} in terms of the determination of "subspace" $\mat{Q}_{i,k}$. Therefore, the equivalence (b) in terms of both "power" and "subspace" is verified since both IA condition and transmit power constraints are jointly considered in the proposed IA transceiver design. The equivalence (c) is based on the definitions $\mat{V}_{i,k}^{in} \triangleq \mat{Q}_{i,k}\tilde{\mat{X}}_{i}, \mat{V}_{i,k}^{out} \triangleq \tilde{\mat{X}}_{i}^{-1}\mat{R}_{i,k}$ and the power constraints
\begin{align}
\trace(\mat{R}_{i,k}\mat{R}_{i,k}^H) 
&= \trace(\mat{Q}_{i,k}\mat{R}_{i,k}\mat{R}_{i,k}^H\mat{Q}_{i,k}^H)   \nonumber \\
&=  \trace(\mat{V}_{i,k}^{in}\tilde{\mat{X}}_{i}^{-1}\mat{R}_{i,k}\mat{R}_{i,k}^H(\tilde{\mat{X}}_{i}^{-1})^H\mat{Q}_{i,k}^H)  \nonumber \\
&= \trace(\mat{V}_{i,k}^{in}\mat{V}_{i,k}^{out}\mat{V}_{i,k}^{out,H}\mat{V}_{i,k}^{in,H}). \nonumber
\end{align}

Due to $\mat{V}_{i,k}^{in}\mat{V}_{i,k}^{out} = \mat{Q}_{i,k}\mat{R}_{i,k} = \mat{V}_{i,k}$, it is equivalent to determine $\mat{V}_{i,k}$ via determining $\mat{V}_{i,k}^{in}$ and $\mat{V}_{i,k}^{out}$ based on the equivalence $(c)$ in place of determining $\mat{Q}_{i,k}$ and $\mat{R}_{i,k}$ based on the equivalence $(a)$.

Therefore, the proposed  restriction-and-relaxation two stages -- first design $\{\mat{V}_{i,k}^{in}\}$ based on $\mat{H}_{1,k}^{k+1}\mat{V}_{1,k}^{in} =  \ldots = \mat{H}_{L,k}^{k+1}\mat{V}_{L,k}^{in}$ and then design $\{\mat{V}_{i,k}^{out}\}$ to maximize the achievable rate subject to $\trace(\mat{V}_{i,k}^{in}\mat{V}_{i,k}^{out}\mat{V}_{i,k}^{out,H}\mat{V}_{i,k}^{in, H}) \leq P_{i,k}~\forall i = 1, \ldots, L$ -- is tight. 
\end{IEEEproof}

\section{Proof of Lemma \ref{le:QuanC}} \label{subsec:ProofQuanC}
\begin{IEEEproof}
The quantization $\widehat{\mat{V}}$ can be exactly expressed by the $N$-dimensional full space $\mat{V} \cup  \mat{V}^{\perp}$ as
 \begin{align}
\widehat{\mat{V}} = \mat{\Pi}_{\mat{V}} \widehat{\mat{V}} + \mat{\Pi}_{\mat{V}}^{\perp} \widehat{\mat{V}} = \mat{V}\mat{C}_1 + \mat{V}^{\perp}\mat{C}_2 , \label{eq:Vexpress2} 
\end{align}
where $\mat{C}_1\in \mathbb{C}^{N\times N}$ and $\mat{C}_2 \in \mathbb{C}^{(M-N)\times N}$ in  (\ref{eq:Vexpress2}) denote the components of $\widehat{\mat{V}}$ projected on the $\mat{V}$ and $\mat{V}^{\perp}$, respectively. 
From (\ref{eq:Vexpress2}), it is derived the properties of $\mat{C}_1$ and $\mat{C}_2$ as
\begin{align}
&\widehat{\mat{V}}^H\widehat{\mat{V}} = \mat{I}_N \Rightarrow \mat{C}_1^H\mat{C}_1 + \mat{C}_2^H\mat{C}_2 = \mat{I}_N; \label{eq:Cpro1} \\
&d_c^2(\widehat{\mat{V}}, \mat{V}) = N - \trace(\widehat{\mat{V}}\widehat{\mat{V}}^H\mat{V}\mat{V}^H) \Rightarrow \nonumber \\ 
&~~~~~~~~~~~~~~~~~~~~~~~~~~~~~~~\trace(\mat{C}_2\mat{C}_2^H) =d_c^2(\widehat{\mat{V}}, \mat{V}).  \label{eq:Cpro2}
\end{align}
By the singular-value decomposition (SVD), $\mat{C}_1$ is expressed by $\mat{C}_1 = {\mat{U}}_{\mat{C}_1}{\mat{\Lambda}}_{{\mat{C}_1}}^{1/2}{\mat{V}}_{{{\mat{C}_1}}}^H$ where eigenvalues ${\mat{\Lambda}}_{{\mat{C}_1}} \triangleq \diag\left\{{\lambda_1(\mat{C}_1^H\mat{C}_1)}, \ldots, {\lambda_N(\mat{C}_1^H\mat{C}_1)}\right \}$ satisfy
${\lambda_n(\mat{C}_1^H\mat{C}_1)} \geq 0, \forall n$ subject to $\sum_{n=1}^N {\lambda_n(\mat{C}_1^H\mat{C}_1)} = N-d_c^2(\widehat{\mat{V}}, \mat{V})$ based on  (\ref{eq:Cpro1}) and (\ref{eq:Cpro2}).
From  (\ref{eq:Cpro1}), we further derive $\mat{C}_2^H\mat{C}_2 = \mat{V}_{\mat{C}_1}(\mat{I}_N - \mat{\Lambda}_{\mat{C}_1})\mat{V}_{\mat{C}_1}^H \succeq \mat{0}_N$, which requires $\lambda_n(\mat{C}_1^H\mat{C}_1) \leq1, \forall n$. Therefore, $\mat{C}_2$ can be expressed by
\begin{align}
\mat{C}_2 = 
\widetilde{\mat{U}}
(\mat{I}_N - {\mat{\Lambda}}_{\mat{C}_1})^{1/2} 
{\mat{V}}_{{\mat{C}_1}}^H
\end{align}
where $\widetilde{\mat{U}} \in \mathbb{C}^{(M-N) \times N}$ satisfying $\widetilde{\mat{U}}^H\widetilde{\mat{U}} = \mat{I}_N$ is to select a $N$-dimensional subspace from the $M-N$-dimensional null space $\mathrm{Span}\{\mat{V}^{\perp}\}$.
\end{IEEEproof}

\section{Proof of Propsosition \ref{pr:InterferenceUpper}}\label{subsec:ProofIAupperbound}
\begin{IEEEproof}
Considering $Cell~k' \stackrel{IA}{\longrightarrow} Cell~k$, we have 
\begin{subequations}
\begin{align}
&\mathcal{I}^k  = \mathcal{I}_{k'}^k = \sum_{i=1}^L\mathcal{I}_{k'}^{i,k} \\
&= \sum_{i=1}^L\sum_{j=1}^L\frac{P_{j,k'}}{d_s\sigma_k^2} \trace\left( \hat{\mat{U}}_{i,k}^H\mat{H}_{j,k'}^{k} \hat{\mat{V}}_{j,k'}\hat{\mat{V}}_{j,k'}^H\mat{H}_{j,k'}^{k,H} \hat{\mat{U}}_{i,k}\right) \label{eq:INL2} \\
& \leq L\sum_{j=1}^L\frac{P_{j,k'}}{d_s\sigma_k^2} \trace\left(\hat{\mat{V}}_{j,k'}^H\mat{H}_{j,k'}^{k,H} \mat{\Pi}_{{\mat{H}}_{j,\ell}^{k} \mat{V}_{j,k'}^{in}}^{\perp}  \mat{H}_{j,k'}^{k} \hat{\mat{V}}_{j,k'} \right)   \label{eq:INL3} \\
&= L\sum_{j=1}^L \frac{P_{j,\ell}}{\sigma_k^2 d_s}  \trace(\mat{S}_{j,k'}^{H}\mat{\Omega}_{j,k'}^k \mat{S}_{j,k'} \mat{\Sigma}_{j,k'})  \label{eq:INL6} \\
&\leq L\sum_{j=1}^L \frac{P_{j,\ell}}{\sigma_k^2 d_s} \sum_{d=1}^{d_s} \lambda_d(\mat{\Omega}_{j,k'}^k) \beta_{j,k'}^d  \label{eq:INL7} \\
&\leq L\sum_{j=1}^L \frac{P_{j,\ell}}{\sigma_k^2 d_s} \lambda_1({\mat{\Omega}_{j,k'}^k}) \sum_{d=1}^{d_s}  \beta_{j,k'}^d  \label{eq:INL8} \\
&= L\sum_{j=1}^L \frac{P_{j,\ell}}{\sigma_k^2 d_s} \lambda_1({\mat{\Omega}_{j,k'}^k}) d_c^2(\widehat{{\mat{V}}}_{j,k'}, \overrightarrow{\mat{V}}_{j,k'})  \label{eq:INL9} \\
& \leq c(N_U, d_s)L\sum_{j=1}^L \frac{P_{j,k'}}{\sigma_k^2 d_s}  \lambda_1(\mat{\Omega}_{j,k'}^{k})  2^{-\frac{B_{j,k'}}{ds(N_U-d_s)}},
\label{eq:INL10}
\end{align}
\end{subequations}
where \eqref{eq:INL2} is based on the definition of $\mathcal{I}_{k'}^{i,k}$ in \eqref{eq:ResiInF1}, and (\ref{eq:INL3}) is derived based on the definition of $\hat{\mat{U}}_{i,k}$ in \eqref{eq: Uinhat}, the inequality of $||\mat{\Pi}_{[{\mat{Y}}_1, {\mat{Y}}_2]}^{\perp}\mat{Y}_3||_{F}^2 \leq ||\mat{\Pi}_{[{\mat{Y}}_1]}^{\perp}\mat{Y}_3||_{F}^2$ and $\mat{\Pi}_{{\mat{H}}_{1,k'}^{k} \mat{V}_{1,k'}^{in}}^{\perp} = \ldots =  \mat{\Pi}_{{\mat{H}}_{L,k'}^{k} \mat{V}_{L,k'}^{in}}^{\perp}$.
Plugging  \eqref{eq: QuanExpress2} into (\ref{eq:INL3}) and removing the zero-valued terms and based on the definition \eqref{eq:Omegadef} yield (\ref{eq:INL6}), where $\mat{S}_{j,k'} \in \mathbb{C}^{(N_U-d_s) \times d_s}$ satisfies $\mat{S}_{j,k'}^H \mat{S}_{j,k'} = \mat{I}_{d_s}$ and $\mat{\Sigma}_{j,k'} = \diag\{\beta_{j,k'}^1, \ldots, \beta_{j,k'}^{d_s}\}, \forall j$ is with $\beta_{j,k'}^d \in (0,1), \forall d$ and $\sum_{d=1}^{d_s} \beta_{j,k'}^d = d_c^2(\widehat{{\mat{V}}}_{j,k'}, \overrightarrow{\mat{V}}_{j,k'})$. 
The upper bound (\ref{eq:INL7}) is achieved when the truncated unitary matrix $\mat{S}_{j,k'}$ is the eigen-subspace of the matrix $\mat{\Omega}_{j,k'}^k$ associated with the $d_s$ largest eigenvalues $\lambda_1(\mat{\Omega}_{j,k'}^k), \ldots, \lambda_{d_s}(\mat{\Omega}_{j,k'}^k)$. 
 (\ref{eq:INL10}) is derived by the quantization distortion upper bound \eqref{eq:Dis2}.
\end{IEEEproof}

\bibliographystyle{IEEEbib}
\bibliography{thesis2}

\begin{IEEEbiography}[{\includegraphics[width=1in,height=1.25in,clip,keepaspectratio]{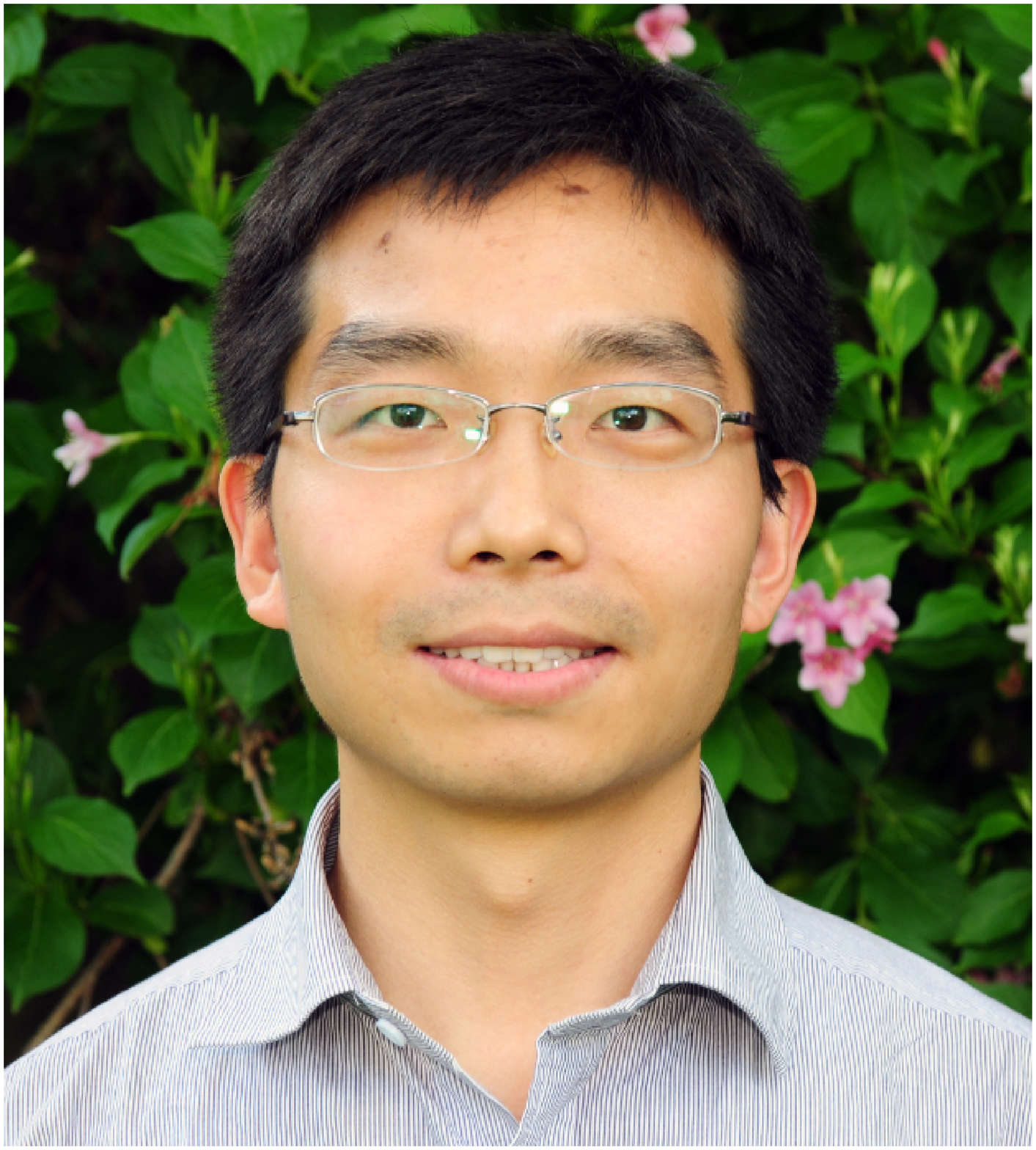}}]{Pan Cao} (S'12 -- M'15) received the B.S. degree in Mechano-Electronic Engineering and the M.S. degree in Information and Signal Processing from Xidian University, P.R. China in 2008 and 2011, respectively, and the Doktor-Ingenieur (Ph.D.) degree in Electrical Engineering from the Technische Universit\"at Dresden, Germany in 2015. Since March 2015, he works as a Postdoctoral Research Associate in the Institutes of Digital Communications at The University of Edinburgh, UK, supported by the EPSRC project SERAN (Seamless and Efficient Wireless Access for Future Radio Networks). 

His current research focuses on designing the novel architectures and algorithms for future wireless communication networks, e.g., dense networks, large scale antenna array and millimeter-wave systems, by optimization techniques and game theory. He received the Best Student Paper Award of the 13th IEEE International Workshop on Signal Processing Advances in Wireless Communications (SPAWC), Cesme, Turkey in 2012, and the Qualcomm Innovation Fellowship (QInF) Award (one of three winners in continent of Europe) in 2013.
\end{IEEEbiography}

\begin{IEEEbiography}[{\includegraphics[width=1in,height=1.25in,clip,keepaspectratio]{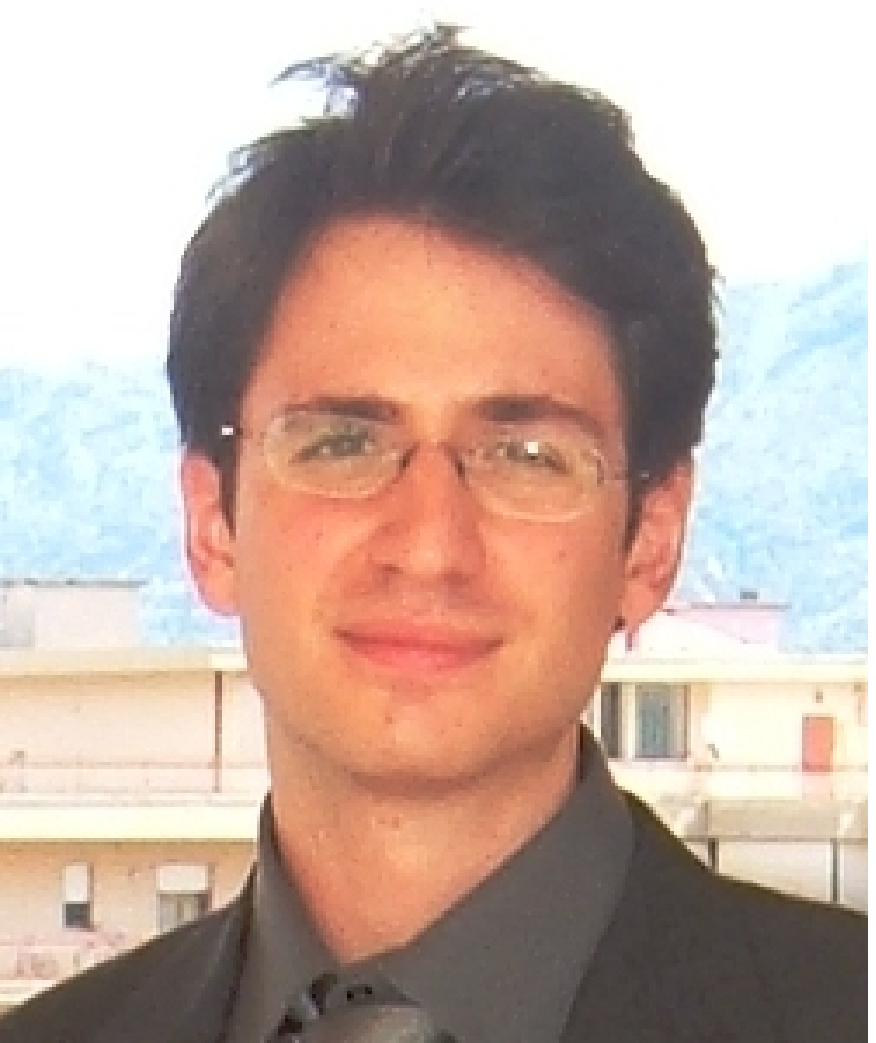}}]{Alessio Zappone} (S'08 -- M'11) is a research associate at the Technische Universit\"at Dresden, Dresden, Germany. Alessio received his M.Sc. and Ph.D. both from the University of Cassino and Southern Lazio. Afterwards, he worked with Consorzio Nazionale Interuniversitario per le Telecomunicazioni (CNIT) in the framework of the FP7 EU-funded project TREND, which focused on energy efficiency in communication networks. Since October 2012, Alessio is the project leader of the project CEMRIN on energy-efficient resource allocation in wireless networks, funded by the German research foundation (DFG).

Alessio's research interests lie in the area of communication theory and signal processing, with main focus on optimization techniques for resource allocation and energy efficiency maximization. He held several research appointments at TU Dresden, Politecnico di Torino, Supélec - Alcatel-Lucent Chair on Flexible Radio, and University of Naples Federico II. He was the recipient of a Newcom\# mobility grant in 2014. 
\end{IEEEbiography}

\begin{IEEEbiography}[{\includegraphics[width=1in,height=1.25in,clip,keepaspectratio]{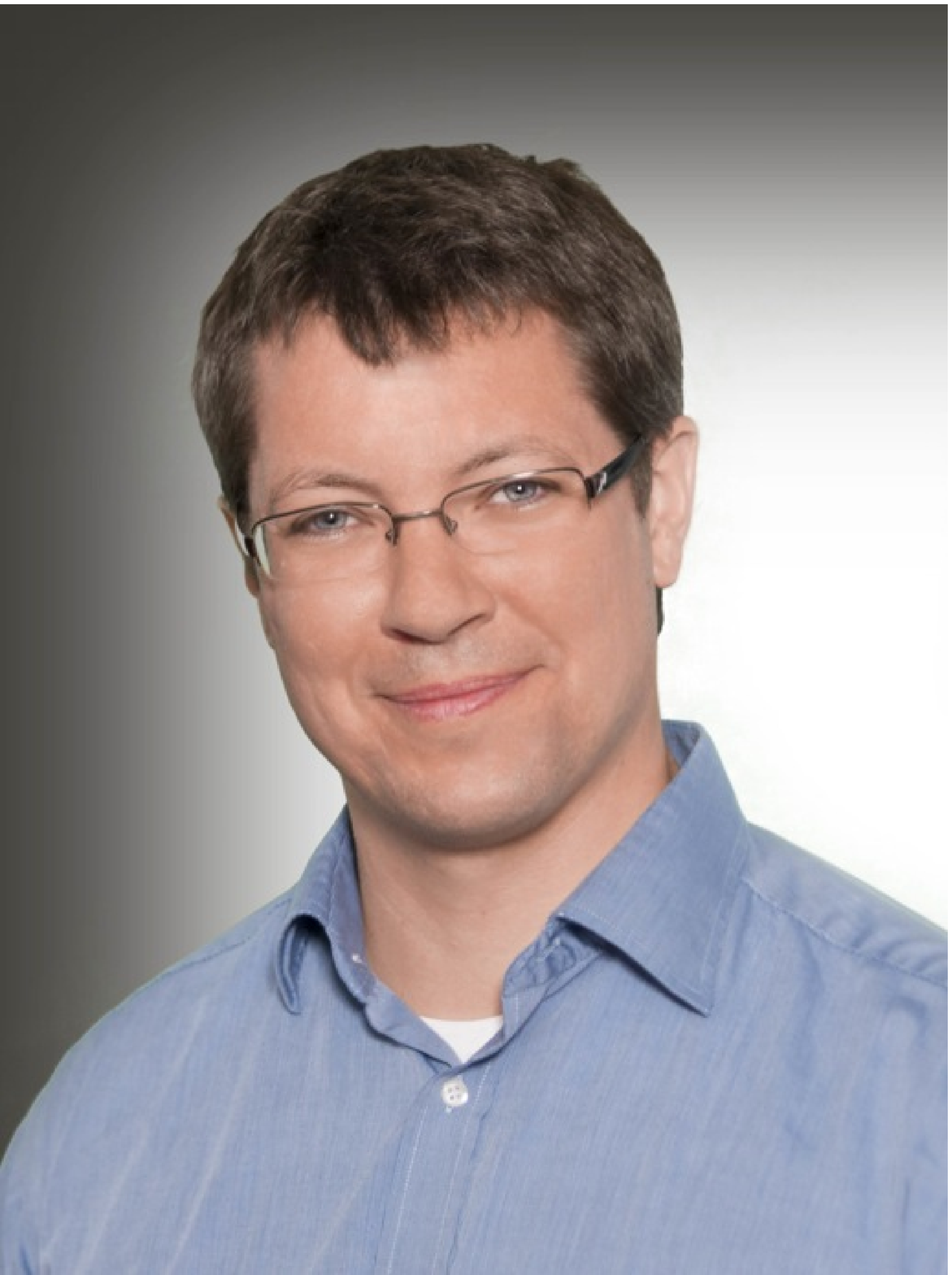}}]{Eduard
Jorswieck} (S'01 -- M'03 -- SM'08) received the Diplom-Ingenieur (M.S.)
degree and Doktor-Ingenieur (Ph.D.) degree, both in electrical
engineering and computer science, from the Technische Universit\"at
Berlin, Germany, in 2000 and 2004, respectively. He was with the
Broadband Mobile Communication Networks Department, Fraunhofer Institute
for Telecommunications, Heinrich-Hertz-Institut, Berlin, from 2000 to
2008. From 2005 to 2008, he was a Lecturer with the
Technische Universität Berlin. From 2006 to 2008,
he was with the Department of Signals, Sensors and Systems, Royal
Institute of Technology, as a Post-Doctoral Researcher and an Assistant
Professor. Since 2008, he has been the Head of the Chair of
Communications Theory and a Full Professor with the Technische
Universit\"at Dresden, Germany. He is principal investigator in
the excellence cluster center for Advancing Electronics Dresden (cfAED)
and founding member of the 5G lab Germany (5Glab.de).

His main research interests are in the area of signal
processing for communications and networks, applied information theory,
and communications theory. He has authored over 80 journal papers, 8
book chapters, some 225 conference papers and 3 monographs on these
research topics. Eduard was a co-recipient of the IEEE Signal Processing
Society Best Paper Award in 2006 and co-authored papers that won the
Best Paper or Best Student Paper Awards at IEEE WPMC 2002, Chinacom
2010, IEEE CAMSAP 2011, IEEE SPAWC 2012, and IEEE WCSP 2012.

Dr. Jorswieck was a member of the IEEE SPCOM Technical Committee
(2008-2013), and has been a member of the IEEE SAM Technical Committee
since 2015. Since 2011, he has been an Associate Editor of the IEEE
TRANSACTIONS ON SIGNAL PROCESSING. Since 2008, continuing until 2011, he
has served as an Associate Editor of the IEEE SIGNAL PROCESSING LETTERS,
and until 2013, as a Senior Associate Editor. Since 2013, he has served
as an Editor of the IEEE TRANSACTIONS ON WIRELESS COMMUNICATIONS.
\end{IEEEbiography}

\ifCLASSOPTIONcaptionsoff
  \newpage
\fi

\end{document}